\documentclass[preprint,12pt,showpacs,aps]{revtex4}
\usepackage{amsfonts}
\usepackage{mathrsfs}
\usepackage{graphicx}
\usepackage{amssymb}
\usepackage{color}
\usepackage{amsmath}
\begin{document}

\title{Non-Gaussian features from the inverse volume corrections in loop quantum cosmology}
\author{Li-Fang Li$^{1,}$\footnote{Email: lilf@itp.ac.cn}}

\author{Rong-Gen Cai$^{1,}$\footnote{Email: cairg@itp.ac.cn}}

\author{Zong-Kuan Guo$^{1,}$\footnote{Email: guozk@itp.ac.cn}}

\author{Bin Hu$^{2,}$\footnote{Email: hubinitp@gmail.com}}
\affiliation{$^{1}$State Key Laboratory of Theoretical Physics,
Institute of Theoretical Physics, Chinese Academy of
Sciences, P.O. Box 2735, Beijing 100190, China}
\affiliation{$^{2}$ INFN, Sezione di Padova, via Marzolo 8, I-35131 Padova, Italy}

\begin{abstract}
In this paper we study the non-Gaussian features of the primordial
fluctuations in loop quantum cosmology with the inverse volume
corrections. The detailed analysis is performed in the single field
slow-roll inflationary models. However, our results reflect the
universal characteristics of bispectrum in loop quantum cosmology.
The main corrections to the scalar bispectrum come from two aspects:
one is the modifications to the standard Bunch-Davies vacuum, the
other is the corrections to the background dependent variables, such
as slow-roll parameters. Our calculations show that the loop quantum
corrections make $f_{{\rm NL}}$ of the inflationary models increase
$0.1\%$. Moreover, we find that two new shapes of non-Gaussian
signal arise, which we name $\mathcal F_{1}$ and $\mathcal F_{2}$.
The former gives a unique loop quantum feature which is less
correlated with the local, equilateral and single types, while the
latter is highly correlated with the local one.

\end{abstract}


\maketitle

\section{Introduction}

As a non-perturbative and background-independent theory, Loop
Quantum Gravity (LQG) \cite{rovelli04,ashtekar04,thiemann07} has
achieved great successes in past years: derivations of the quantized
area and volume operators
\cite{rovelli95,ashtekar97,ashtekar98,thiemann98a}, calculations of
black holes entropy \cite{rovelli96} and Loop Quantum Cosmology(LQC)
\cite{bojowald08a}, {\it etc.} And the nonperturbative quantization
procedure of LQG is also valid for a more general class of
four-dimensional metric theories of
gravity~\cite{zhang11a,zhang11b,zhang11c}. As an example of LQG, LQC
gives a quantization scheme of LQG for a symmetry-reduced model in
the homogeneous and isotropic Friedmann-Lema\^ itre-Robertson-Walker
universe. The discrete spacetime geometry in LQC scenarios predicts
a non-singular bouncing universe in some simplified models, which
satisfies most of the astronomical and cosmological observational
constraints. Although the quantum correction effects are being
diluted with the expansion of our universe, it remains present in a
weaker form, especially on/near the super-horizon scales.

Recently, the gauge invariant cosmological perturbation theory has
been systematically constructed in
\cite{bojowald06,bojowald08,bojowald09} for inverse volume
corrections and in \cite{wu10,Cailleteau11} for holonomy
corrections. Some relevant applications have been considered
in~\cite{bojowald11a,bojowald11b,bojowald11c,zhu11}. The inverse
volume corrections and the holonomy corrections are two main quantum
corrections in LQC. The inverse volume corrections come from the
quantization of the inverse of the volume operator in LQG. The
inverse volumes exist in the Hamiltonian constraint of gravity and
the usual matter Hamiltonian, especially in the kinematic terms.
Since the volume can be taken the value zero, there does not exist
well defined inverses of the volume operator. Fortunately, with the
Thiemann trick~\cite{thiemann98}, we can construct well defined
inverse volume operators, which bring the quantum corrections. While
the holonomy corrections arise from the loop quantization based on
the holonomies instead of the direct connection. The holonomy
corrections become important when the energy scale of our Universe
approaches the Planck one. Both the modifications to the scalar and
tensor primordial power spectra from the inverse volume corrections
are carefully investigated by the authors of \cite{bojowald11a}.
Their results show that the inverse volume corrections could give
rise to the enhancement of the power spectra on the large scales,
i.e., a red-tilt one. However, some other mechanisms such as the
non-commutative geometry or string
theory~\cite{tsujikawa03,piao04,calcagni04}, could also lead to
similar features. Therefore to seek for signature of loop quantum
cosmology, the study for the non-Gaussianity features in loop
quantum cosmology is necessary.

Because the primordial non-Gaussianities are quite helpful to
distinguish inflationary models, so far a lot of papers have been
devoted to studying the non-Gaussianities in different inflation
models, see the relevant references in the nice reviews
\cite{Bartolo:2004if,Chen:2010xk}. Inspired by the studies of
\cite{bojowald11a}, in this paper we mainly consider the
non-Gaussianities from the inverse volume corrections in LQC. The
reason for consider inverse volume corrections only is as following.
We denote $\delta_{\rm inv}$ as the correction term coming from the
inverse volume operator and $\delta_{{\rm hol}}$ as the correction
term from the holonomy corrections. We can estimate the inverse
volume correction as\cite{bojowald11c}
\begin{equation}\label{compare}
\delta_{{\rm inv}}\sim\left(\frac{8\pi}{3}\frac{\rho}{\rho_{{\rm
Pl}}}\delta_{{\rm hol}}^{-1}\right)^2\;,
\end{equation}
where the Planck density $\rho_{{\rm Pl}}$ is assumed as the quantum
gravity scale. From the above expression, we can see that the
inverse volume corrections behave very differently from what is
normally expected for quantum gravity. For low densities, the
holonomy corrections is small, but the inverse volume one may still
be large because they are magnified by the inverse of $\delta_{{\rm
hol}}$. For an example, the small holonomy corrections of size
$\delta_{{\rm hol}}<10^{-6}$ then requires the inverse volume
corrections larger than $\delta_{{\rm inv}}>10^{-6}$ even at scale
$\rho\approx10^{-9}\rho_{\rm Pl}$. These novel features make the
investigations on the inverse volume corrections more interesting
than the former at sub-Planckian inflationary scales. So we only
consider the inverse volume correction in this work.

Explicitly, in the perturbation theory in LQC, the inverse volume
operator can be captured by a correction function such as
$\bar{\alpha}\simeq1+\alpha_0\delta_{\rm inv}\simeq
1+\alpha_0(a_{\rm inv}/a)^{\sigma}$, where $a$ is the scale factor
of the FLRW universe and $a_{\rm inv}$ is introduced to describe the
characteristic scale of the inverse volume correction, which is not
the Planck one in general. When $a_{\rm inv}/{a}\ll 1$, we can
ignore the correction term. However, if $a_{inv}/a \lesssim 1$
during inflation, one cannot neglect inverse volume corrections. In
this case, the correction approximates $\alpha_0\delta_{\rm
inv}(k)\approx\delta(k_0)(k_0/k)^{\sigma}$, where $k$ and $k_0$ are,
respectively, the considered perturbation wave number and some
characteristic number involved in the inverse volume correction. In
addition, many works about LQC imply that $\sigma\in[0,6]$
\cite{bojowald11b}. From the form of the inverse volume correction,
we can see that a small sigma corresponds to a small the inverse
volume correct, vice versa. Therefore, as an example, following
\cite{bojowald11b}, we take $\sigma=2$ in this paper. For other
$\sigma$ values the behavior will be similar. Furthermore, in terms
of spherical multiples the wave number could be expressed as
$k\approx10^{-4}hl$, with $h$ representing for the reduced Hubble
parameter $h=0.7$ and $l$ for the spherical multiples. In the
typical linear regime of Cosmic Microwave Background (CMB), the
multiples $l$ range in $2<l<1000$ or more. Given all the mentioned
observations, we expect some new features in the non-Gaussianities
will arise. The purpose of this work is to investigate the
characterized sizes and shapes of bispectra in LQC scenarios.

Note that the bispectra for the single field slow-roll inflationary
model have been calculated in the
papers~\cite{arXiv:0209156,maldacena03,Seery:2005wm,Chen:2006nt,Chen:2006xjb}.
For simplicity and comparison, we study the simplest single field
slow-roll inflationary model in LQC. Our results show that the
quantum corrections mainly come from the third order interaction
Hamiltonian, the corrected vacuum state and the corrections to the
slow-roll parameters. Our paper is organized as follows. In
Sec.~\ref{sec:review}, the canonical formulism and the slow-roll
inflationary model in LQC scenarios are briefly reviewed. In
Sec.~\ref{sec:power}, we study the power spectrum in LQC and recover
the previous results. The effect of the inverse volume corrections
on the non-Gaussianity in LQC is investigated in
Sec.~\ref{sec:bispectrum}. The detailed analysis for the sizes,
shapes and shape correlations is presented in
Sec.~\ref{sec:bispectrum}and Sec.~\ref{sec:shape}, respectively.
Throughout this paper we set $8\pi \gamma G=1$ and Einstein's summing
convention is always adopted.

\section{Review of Loop Quantum Cosmology}
\label{sec:review} The framework of LQG/LQC will be briefly
presented in this section. Firstly, we discuss the canonical
formalism in LQG, and then introduce the dynamics of slow-roll
inflationary models in LQC scenarios.

\subsection{The canonical formalism in loop quantum gravity}
In the framework of LQG~\cite{rovelli04,thiemann07}, the spatial
metric as a canonical field is replaced by the densitized triad
$E^a_i$, defined as
\begin{eqnarray}
E^a_i:=|\det (e^j_b)|e^a_i,
\end{eqnarray}
where $e^a_i$ is the inverse of the cotriad $e^i_a$ related to the
spatial metric by $q_{ab}=e^i_ae^i_b$. The canonically conjugate
variable to the densitized triad is the Ashtekar-Barbero connection
$A^i_a:=\Gamma^i_a+ \gamma K^i_a$, where $K^i_a$ is the extrinsic
curvature and $\gamma\approx0.274$ is the Barbero-Immirzi
parameter\cite{domagala04,meissner04}. The densitized triad $E^a_i$
and the Ashtekar-Barbero connection $A^i_a$ satisfy the following
commutator relation
\begin{eqnarray}
\{A^i_a(x), E^b_j(y)\}=\delta^b_a\delta^i_j\delta^3(x,y).
\end{eqnarray}
The spin connection $\Gamma^i_a$ is defined such that it leaves the
triad covariantly constant and has the explicit form
\begin{eqnarray}
\Gamma^i_a=-\epsilon^{ijk}e^b_j(\partial_{[a}e^l_{b]}\delta_{lk}+\frac{1}{2}e^c_ke^l_a\partial_{[c}e^m_{b]}\delta_{lm}).
\end{eqnarray}
In the new Ashtekar variables, the Einstein-Hilbert action can be
expressed in the canonical form
\begin{eqnarray}
S_{EH}=\int dt\Big[\int_{\Sigma} d^3x
\dot{K^i_a}E^a_i-\mathscr{D}_{{\rm grav}}[N^a]-\mathscr{H}_{{\rm
grav}}[N]-\mathscr{G}_{{\rm grav}}[\Lambda^i]\Big]
\end{eqnarray}
where the Diffeomorphism constraint is
\begin{eqnarray}
\mathscr{D}_{{\rm grav}}[N^a]=\frac{1}{\gamma}\int_{\Sigma}d^3
xN^a\Big[(\partial_aA^j_b-\partial_b
A^j_a)E^b_j-A^j_a\partial_bE^b_j\Big].
\end{eqnarray}
And correspondingly the Hamiltonian constraint can be expressed as
\begin{eqnarray}
\mathscr{H}_{{\rm grav}}[N]=\frac{1}{2}\int_{\Sigma}d^3x N
\epsilon_i^{jk}\frac{E^c_jE^d_k}{\sqrt{|\det
E|}}\Big[2\partial_c\Gamma^i_d+\epsilon^i_{mn}(\Gamma^m_c\Gamma^n_d-K^m_cK^n_d)\Big].
\end{eqnarray}
The Gaussian constraint is
\begin{eqnarray}
\mathscr{G}_{{\rm
grav}}[\Lambda^i]=\int_{\Sigma}d^3x\Lambda^i(\partial_aE^a_i+\epsilon^{k}_{ij}\Gamma^j_aE^a_k+\epsilon^{k}_{ij}K^j_aE^a_k),
\end{eqnarray}
which can be solved through standard procedure~\cite{bojowald09}.
Thus, solutions for the scalar mode perturbations are completely
determined by the Hamiltonian constraint and the Diffeomorphism
constraint.

\subsection{Slow-roll inflationary models}
In this subsection, we shortly review the inflationary dynamics of
the Friedmann-Lema\^ itre-Robertson-Walker Universe in LQC
scenarios. The modified Friedmann equation, Raychaudhuri equation
and Klein-Gordon equation are respectively \cite{bojowald09}
\begin{eqnarray}
{\mathcal H}^2&=&\frac{1}{3}\bar{\alpha}\left(\frac{\bar
{\varphi}^{'2}}{2\bar\nu}+\bar{p}V(\bar\varphi)\right)\label{FriBG},\\
{\mathcal H}'&=&{\mathcal H}^2\left(1+\frac{\bar\alpha_{,\bar{\rm
p}} \bar
p}{\bar\alpha}\right)-\frac{1}{2}\frac{\bar\alpha}{\bar\nu}{\bar\varphi}^{'2}\left(1-\frac
{\bar\nu_{,\bar{\rm p}}
\bar p}{3\bar\nu}\right)\label{RayBG},\\
{\bar\varphi}''&+&2 {\mathcal
H^2}{\bar\varphi}'\left(1-\frac{\bar\nu_{,\bar{\rm p}}{\bar
p}}{\bar\nu} \right)+\bar\nu \bar p
V_{,\varphi}(\bar\varphi)=0\label{KGBG},
\end{eqnarray}
where $\mathcal H=\frac{\bar{p}'}{2\bar{p}}$, $\bar{\nu}_{,\bar{\rm
p}}\equiv d\bar\nu/d \bar{p}$, $\bar{p}\equiv a^2$ and a prime
represents the derivative with respect to the conformal time.
$\bar{\alpha}$ and $\bar{\nu}$ are the correction functions for the
inverse volumes and they read $\bar{\alpha}\approx
1+\alpha_0\delta_{\rm inv}$, $\bar{\nu} \approx 1+\nu_0\delta_{\rm
inv}$.

Following~\cite{bojowald11b,bojowald11c}, the slow-roll parameter
$\epsilon$ can be straightforwardly calculated as
\begin{eqnarray}
\epsilon&=&1-\frac{\mathcal{H}'}{\mathcal{H}^2}\;,\nonumber\\
&=&\epsilon_0(1+\gamma_{\epsilon}\delta_{{\rm
inv}})\;,\label{slowrll1}
\end{eqnarray}
where $\epsilon_0$ denotes for the usual slow-roll parameter. And
the explicit form of coupling constant $\gamma_{\epsilon}$ reads
\begin{equation}
\gamma_{\epsilon}=-\delta_0\left[\frac{\sigma\alpha_0}{2\epsilon_0}
+\alpha_0+\nu_0\left(\frac{\sigma}{6}-1\right)\right]\;.
\end{equation}
Typically we set $\alpha_0=0.06$, $\nu_0=0.17$ and $\epsilon_0=0.01$
in this paper~\cite{bojowald11b}. $\delta_0$ is determined by
quantum correction and the analysis after Eq.(\ref{F}) implies that
it takes $\mathcal O(10^{-3})$. As mentioned in Introduction we set
$\sigma\in[0,6]$ in this work. From above expression we can easily
estimate that $\gamma_{\epsilon}$ is of the order $\mathcal
O(10^{-3})$. Formally we can also express the another slow-roll
parameter $\eta$ as
\begin{eqnarray}
\eta&=&1-\frac{\varphi''}{\mathcal H\varphi'}\;,\nonumber\\
&=&\eta_0(1+\gamma_{\eta}\delta_{{\rm inv}})\;,\label{slowrll2}
\end{eqnarray}
where $\eta_0$ denotes for the usual slow-roll parameter as
$\epsilon_0$.

As in the usual situation of the single field inflation model, we
can assume that the two slow-roll parameters, $\epsilon_0$ and
$\eta_0$, take roughly the same order as $10^{-2}$. Thus the typical
values of coupling constant $\gamma_{\eta}$ and $\gamma_{\epsilon}$
are of the order $\mathcal O(10^{-3})$. The terms proportional to
the $\delta_{{\rm inv}}$ represent for the inverse volume
corrections. Here, we should emphasize that the subscript ``inv'' is
introduced to avoid confusion with perturbations, such as
$\delta\varphi$.

\section{power spectrum}
\label{sec:power} In this section, we will firstly review the
formalism of scalar perturbations in LQC; then, derive the second
order Hamiltonian; and finally calculate the primordial power
spectrum in the spatially flat gauge.
\subsection{Formalism on the scalar modes}
Consider the scalar perturbations only, the general form of a
perturbed metric around the isotropic FRW background is
\begin{equation}\label{MetricPert}
d s^2 =
a^2(\tau)\Big\{-(1+2\phi) d \tau^2 +2\partial_a B d \tau d x^a +
\Big[(1-2\psi)\delta_{ab}+2\partial_a\partial_b E\Big]d x^a d x^b\Big\}\,,
\end{equation}
where the scalar factor $a$ is a function of the conformal time
$\tau$, and $(\phi,\psi,E,B)$ are the four scalar metric
perturbations. In the perturbation theory, the triad can be
described by
\begin{eqnarray}
E^a_i=\bar{E^a_i}+\delta E^a_i\label{triad}\;,
\end{eqnarray}
where
\begin{eqnarray}
\bar{E^a_i}=\bar{p}\delta^a_i,\ \ \ \ \delta
E^a_i=-2\bar{p}\psi\delta^a_i+\bar{p}(\delta^a_i\Delta-\partial^a\partial_i
)E.
\end{eqnarray}
The perturbed triad is described by the spatial part of the
perturbed metric $\psi$ and $E$. Here $\Delta$ is the laplace
operator in the flat space. Similarly, the perturbed lapse function
and shift vector can be described by the other two scalar metric
perturbation $\phi$ and $B$ respectively,
\begin{equation}
\delta N=\bar{N}\phi,\ \ \ \ \ \ \ \delta N^a=\partial^a B.
\end{equation}
The extrinsic curvature can be perturbed as
\begin{equation}
K^i_a=\bar{K^i_a}+\delta K^i_a=\bar{k}\delta^i_a+\delta K^i_a.
\end{equation}
For a general triad (\ref{triad}), the linearized spin connection
becomes
\begin{equation}
\delta\Gamma^i_a=\frac{1}{2\bar{p}}\epsilon^{ij}_a\partial_b\delta
E^b_j.
\end{equation}

As described above, the symplectic structure also splits into two
parts, one for the background variables and the other for the
perturbations,
\begin{eqnarray}
\{\bar{k},\bar{p}\}=\frac{1}{3V_0},\ \ \ \ \{\delta K^i_a(x),\delta
E^b_j(y)\}=\delta^3(x,y)\delta^b_a\delta^i_j\;,
\end{eqnarray}
where the background variables are defined by
\begin{equation}
\bar{p}=\frac{1}{3V_0}\int E^a_i\delta^i_a d^3x,\ \ \
\bar{k}=\frac{1}{3V_0}\int K^i_a\delta^a_id^3x\;.
\end{equation}
Here $V_0$ is some artificial finite volume.

In this paper, the matter part is represented by a scalar field
$\varphi$. Similarly, we split the field $\varphi$ and its conjugate
momentum $\pi$ into homogeneous part and inhomogeneous one as well
\begin{eqnarray}
\varphi=\bar{\varphi}+\delta \varphi,\ \ \pi=\bar{\pi}+\delta \pi\;.
\end{eqnarray}
Hence, the basic Poisson brackets are reduced into
\begin{eqnarray}
\{\bar{\varphi},\bar{\pi}\}=\frac{1}{V_0}, \ \ \
\{\delta\varphi(x),\delta\pi(y)\}=\delta^3(x-y).
\end{eqnarray}

For simplicity, we introduce the LQC formalism with finite cell
$V_0$ rather than the whole $R^3$ region in above description. But
the unphysical feature of $V_0$ can be remedied by lattice
refinement model \cite{sakellariadou09}. Since our following
calculation only involves $\delta_{\text{inv}}$ we adopt the lattice
refinement parametrization procedure in \cite{bojowald11b} to
eliminate the effect of artificial volume $V_0$.
\subsection{The second order Hamiltonian}

According to \cite{bojowald09}, the quantum corrected second order
Hamiltonian constraint can conveniently be written as
\begin{eqnarray}
\mathscr{H}^{(2)}&=&\mathscr{H}_{\rm
grav}^{(2)}[\bar{N}]+\mathscr{H}_{\rm grav}^{(2)}[\delta
N]+\mathscr{H}_{\rm matter}^{(2)}[\bar{N}]+\mathscr{H}_{\rm
matter}^{(2)}[\delta
N]\nonumber\\
&=&\frac{1}{2}\int_{\Sigma} d^3x\bar{N}\bar{\alpha}\mathfrak{H}_{\rm
grav}^{(2)}+\frac{1}{2}\int_{\Sigma} d^3x\delta N
\bar{\alpha}\mathfrak{H}_{\rm
grav}^{(1)}+\int_{\Sigma}d^3x\bar{N}\Big[\bar{\nu}
\mathfrak{H}_{\rm \pi}^{(2)}+\bar{\theta}\mathfrak{H}_{\rm \nabla}^{(2)}+\mathfrak{H}_{\rm \varphi}^{(2)}\Big]\nonumber\\
&+&\int_{\Sigma}d^3x\delta N[\bar{\nu}\mathfrak{H}_{\rm
\pi}^{(1)}+\mathfrak{H}_{\rm \varphi}^{(1)}]\;,
\end{eqnarray}
where
\begin{eqnarray}
\mathfrak{H}_{\rm grav}^{(1)} &=& -4(1+f) \bar{k}\sqrt{\bar{p}}
\delta^c_j\delta K_c^j -(1+g)\frac{\bar{k}^2}{\sqrt{\bar{p}}}
\delta_c^j\delta E^c_j +\frac{2}{\sqrt{\bar{p}}}
\partial_c\partial^j\delta E^c_j  ~,
\\
\mathfrak{H}_{\rm grav}^{(2)} &=& \sqrt{\bar{p}} \delta K_c^j\delta
K_d^k\delta^c_k\delta^d_j - \sqrt{\bar{p}} (\delta
K_c^j\delta^c_j)^2 -\frac{2\bar{k}}{\sqrt{\bar{p}}} \delta
E^c_j\delta K_c^j\nonumber
\\
&& \quad -\frac{\bar{k}^2}{2\bar{p}^{3/2}} \delta E^c_j\delta
E^d_k\delta_c^k\delta_d^j +\frac{\bar{k}^2}{4\bar{p}^{3/2}}(\delta
E^c_j\delta_c^j)^2 -(1+h)\frac{\delta^{jk}
}{2\bar{p}^{3/2}}(\partial_c\delta E^c_j) (\partial_d\delta E^d_k)
~,\\
\mathfrak{H}_{\rm \pi}^{(1)} &=& (1+f_1)\frac{\bar{\pi}
\delta{\pi}}{\bar{p}^{3/2}}
-(1+f_2)\frac{\bar{\pi}^2}{2\bar{p}^{3/2}} \frac{\delta_c^j
\delta E^c_j}{2\bar{p}}\;,\\
\mathfrak{H}_{\rm \nabla}^{(1)} &=& 0\;,\\
\mathfrak{H}_{\rm \varphi}^{(1)} &=& \bar{p}^{3/2}\left(
(1+f_3)V_{,\rm \varphi}(\bar{\varphi}) \delta\varphi
+V(\bar{\varphi}) \frac{\delta_c^j \delta
E^c_j}{2\bar{p}}\right)\;,\\
\mathfrak{H}^{(2)}_{\pi}&=&
(1+g_1)\frac{{\delta{\pi}}^2}{2\bar{p}^{3/2}}
-(1+g_2)\frac{\bar{\pi} \delta{\pi}}{\bar{p}^{3/2}} \frac{\delta_c^j
\delta E^c_j}{2\bar{p}}
+\frac{1}{2}\frac{\bar{\pi}^2}{\bar{p}^{3/2}} \left(
(1+g_3)\frac{(\delta_c^j \delta E^c_j)^2}{8\bar{p}^2}
+\frac{\delta_c^k\delta_d^j\delta E^c_j\delta E^d_k}{4\bar{p}^2}
\right)\;,\nonumber\\
\\
\mathfrak{H}^{(2)}_{\nabla}&=&\frac{1}{2}(1+g_5)\sqrt{\bar{p}}\delta^{ab}\partial_a\delta
\varphi
\partial_b\delta \varphi\;,\\
\mathfrak{H}^{(2)}_{\varphi} &=&\bar{p}^{3/2}
\left[(1+g_6)\frac{1}{2} V_{,\varphi\varphi}(\bar{\varphi})
{\delta\varphi}^2 + V_{,\rm \varphi}(\bar{\varphi}) \delta\varphi
\frac{\delta_c^j \delta E^c_j}{2\bar{p}}+ V(\bar{\varphi})\left(
\frac{(\delta_c^j \delta E^c_j)^2}{8\bar{p}^2}
-\frac{\delta_c^k\delta_d^j\delta E^c_j\delta E^d_k}{4\bar{p}^2}
\right)\right] \,,\nonumber\\
\end{eqnarray}
where the definitions of the counterterms can be found in
\cite{bojowald08} or in the Appendix B of \cite{bojowald09}. And the
perturbed second order diffeomorphism constraint can be expressed as
\begin{eqnarray}
\mathscr{D}^{(2)}[\delta N^a]&=&\mathscr{D}^{(2)}_{\rm grav}[\delta
N^a]+\mathscr{D}^{(2)}_{\rm matter}[\delta
N^a]\nonumber\\&=&\int_{\Sigma}d^3x\delta
N^a\Big[\bar{p}\partial_a(\delta^d_k\delta
K^k_d)-\bar{p}(\partial_k\delta
K^k_a)-\bar{k}\delta^k_a(\partial_d\delta
E^d_k)+(\bar{\pi}\partial_a\delta\varphi)\Big]\;.
\end{eqnarray}
Based on this corrected Hamiltonian, we can get the homogeneous and
inhomogeneous part of matter field as
\begin{align}
\bar{\pi}=\bar{\varphi}'\bar{p}/\bar{\nu}, \delta
\pi=\bar{p}\left\{\Big[\delta
\varphi'-\bar{\varphi}'(1+f_1)\phi\Big](1-g_1)+\bar{\varphi}'\frac{\delta
E^a_i\delta^i_a}{2\bar{p}}\right\}/\bar{\nu},
\end{align}
where $f_1$ and $g_1$ are the counterterms.

\subsection{Power spectrum}

Calculations can be simplified greatly in the spatially flat gauge
($\psi=0$, $E=0$), because the perturbed triad vanishes ($\delta
E^a_i=0$) in this gauge. Here we fix the gauge after having put
quantum corrections in Hamiltonian and having checked consistency.
In contrast, in references \cite{gaugeafter}, the authors fixed the
gauge beforehand. We believe our treatment is more consistent. By
solving the constraint equations, the perturbed lapse function and
shift vector read
\begin{eqnarray}
\phi&=&\frac{1}{2}\frac{\bar{\alpha}}{\bar{\nu}}\frac{{\bar{\varphi}}'}{{\mathcal
H}}\frac{1}{1+f}\delta \varphi,\\
\Delta
B&=&-\frac{1}{2}\frac{\bar{\alpha}}{\bar{\nu}}\frac{1}{{\mathcal
H}}\frac{1+f_3}{1+f}\{{\bar{\varphi}'}{\delta
\varphi}'-{\bar{\varphi}}^{'2}(1+f_1)\phi+\bar{\nu}\bar{p}V_{,\varphi}(\bar{\varphi})\delta
\varphi\}-3{\mathcal H}(1+f)\phi\;.
\end{eqnarray}
And the extrinsic curvature is
\begin{eqnarray}
&\bar{\alpha}\delta K^i_a=-\delta^i_a {\mathcal
H}(1+f)\phi-\partial_a\partial^i B,
\end{eqnarray}
where $\bar{\alpha}\bar{k}={\mathcal H}$.

In the spatially flat gauge, the total second order Hamiltonian
becomes
\begin{eqnarray}
\mathscr{H}^{(2)}&=&\int_{\Sigma}d^3x\Big\{\Big[\frac{3\bar{p}\bar{\alpha}}{2\bar{\nu}^2}
-\frac{\bar{\alpha}^2\bar{p}}{4\bar{\nu}^3}
\frac{{\bar{\varphi}}^{'4}}{{\mathcal H}^2}(1-g_1+2f_1-2f)
\nonumber\\
&+&\frac{(1+g_6)}{2\bar{p}^2}V_{,\varphi\varphi}(\bar{\varphi})
+\frac{\bar{\alpha}\bar{p}^2}{\bar{\nu}}\frac{{\bar{\varphi}}'}{{\mathcal
H}}V_{,\varphi}(\bar{\varphi})(1+f_3-f)\Big]\delta\varphi^2\nonumber\\
&+&\frac{\bar{p}}{2\bar{\nu}}(1-g_1)\delta{\varphi'}^2
+\frac{\bar{p}\bar{\theta}}{2}(1+g_5)\delta^{ab}\partial_a\delta\varphi\partial_b\delta\varphi\Big\}\;,
\end{eqnarray}
where $\bar{\theta}$ is also a correction function for the inverse
volume and $\bar{\alpha}^2=\bar{\nu}\bar{\theta}$. In this gauge,
the dynamical inflaton perturbation $\delta\varphi$ coincides with
the Sasaki-Mukhanov variable $u=z\zeta$, with
\begin{eqnarray}
\label{z} z=\frac{\varphi'}{\mathcal
H}\left[1+\left(\frac{\alpha_0}{2}-\nu_0\right)\delta_{{\rm
inv}}\right]\;.
\end{eqnarray}
Then, one can derive the Mukhanov equation \cite{bojowald11b}
\begin{eqnarray}
u''-(c_s^2\Delta+\frac{z''}{z})u=0\;,
\end{eqnarray}
where $c_s$ is the propagation speed of the perturbation. The
solution of the above equation is \cite{bojowald11b}
\begin{eqnarray}
\label{sol_vacu}
u(k,\tau)&=&\frac{H}{\sqrt{2k^3}}e^{-ik\tau}\Big[1+ik\tau-\frac{\chi}{2(\sigma+1)}(1+ik\tau)\delta_{{\rm inv}}\Big]\nonumber\;,\\
&=&\frac{H}{\sqrt{2k^3}}e^{-ik\tau}\left[F\left(\frac{k_0}{k}\right)+ikF\left(\frac{k_0}{k}\right)
\tau+\mathcal{O}(k^2\tau^2)\right]\;,
\end{eqnarray}
where
\begin{equation}
\label{F} F\left(\frac{k_0}{k}\right)=\left[1-\frac{\chi}{2(\sigma
+1)}\delta_{\rm
inv}\right]=\Big[1+C\left(\frac{k_0}{k}\right)^{\sigma}\Big]\;,
\end{equation}
where $\chi=\sigma\nu_0(1+\sigma/6)/3+\alpha_0(5-\sigma/3)/2$,
$C=-\frac{\delta_0\chi}{2(\sigma+1)}$ and
$\delta_0=\delta(k_0)/\alpha_0$. The latter variable $\delta(k_0)$
is constrained by the cosmic observational data~\cite{bojowald11c},
such as Cosmic Microwave Background and Large Scale Structures. For
the specific inflationary models with a quadratic potential and
$\sigma=2$, $\delta(k_0)\sim \mathcal O(10^{-5})$. In this paper we
take $\alpha_0\sim\mathcal O(10^{-2})$, i.e., the variable
$\delta_0$ is of the order $\mathcal O(10^{-3})$. Moreover, Eq.
(\ref{sol_vacu}) tells us that, on the one hand the inverse volume
corrections become important for long wave modes with $k\ll k_0$; on
the other hand long wave modes cross horizon earlier than the short
ones. It means that the inverse volume correction will leave more
hints on large scales than small ones. These features are much
different from those of the inflationary models with higher
derivative terms such as K-inflation in Einstein gravity.

Using the canonical quantization, we have
\begin{eqnarray}
\zeta({\vec{k}},\tau)=\zeta^++\zeta^-=\zeta(\vec{k},\tau)a_{\vec{k}}+\zeta^*(\vec{k},\tau)a^{\dag}_{-\vec{k}},
\end{eqnarray}
where $\zeta(\vec{k},\tau)=u(\vec{k},\tau)/z(\vec{k})$. Then the
two-point correlation functions of curvature perturbations can be
calculated straight forwardly as
\begin{eqnarray}
\langle
\zeta_{\vec{k}}\zeta_{\vec{k}'}\rangle=\frac{|u|^2}{z^2}\approx
(2\pi)^3\delta(\vec{k}+\vec{k}')\frac{H^2}{2\epsilon_0
k^3}\{1+\gamma_s \delta_{\rm inv}\}\;,
\end{eqnarray}
with
\begin{eqnarray}
\label{gamma_s}
\gamma_s=\nu_0(1+\sigma/6)+\sigma\alpha_0/2\epsilon-\chi/(\sigma+1)
\end{eqnarray}
where the first and second terms in (\ref{gamma_s}) comes from the $z(\vec{k})$ factor
in the gauge transformations, and the third term attributes to the modifications
of vacuum state (\ref{sol_vacu}).

\begin{widetext}
\begin{figure}
\begin{picture}(500,190)(0,0)
\put(-15,-15)
{
\scalebox{0.3}{\includegraphics{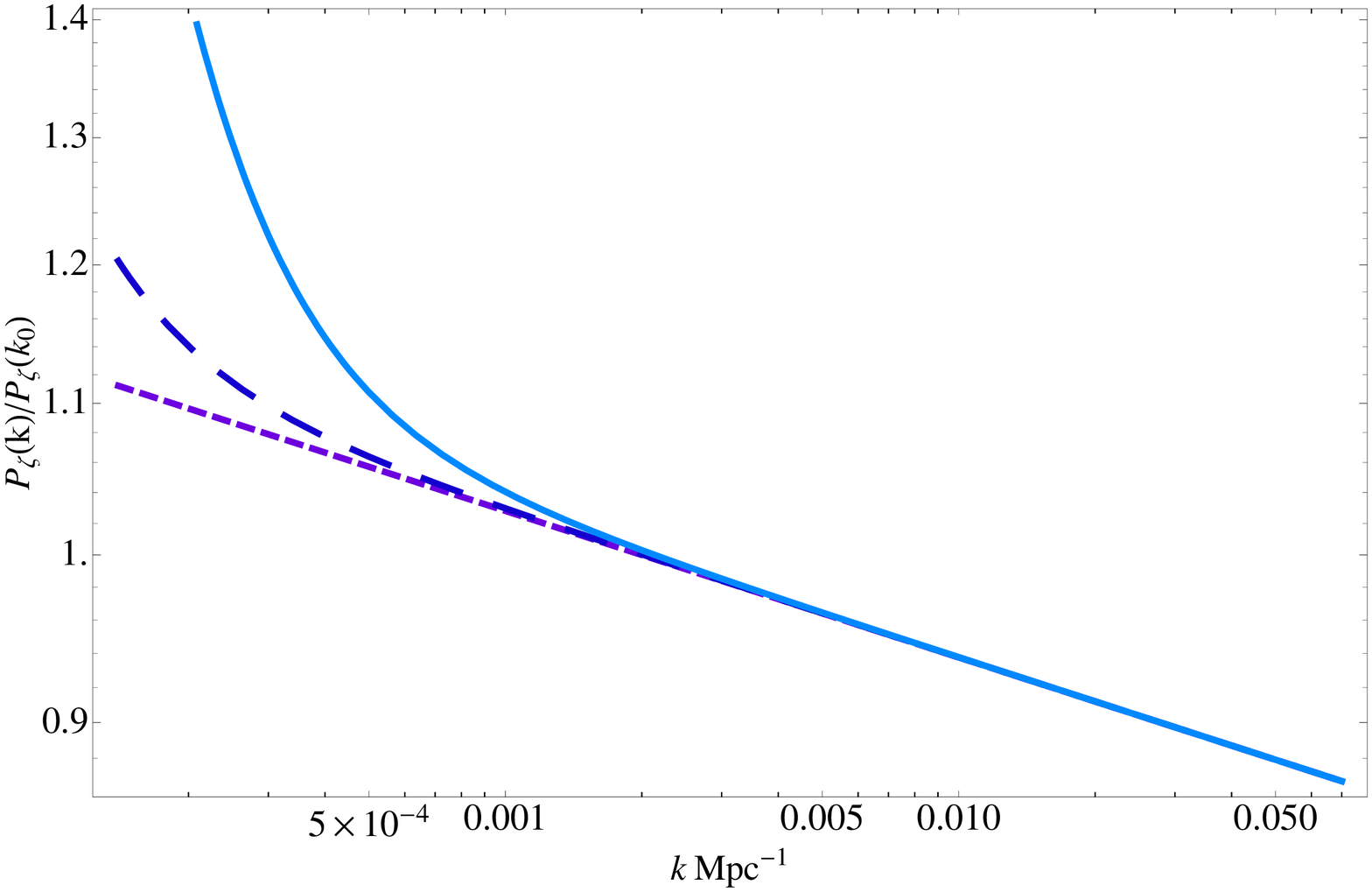}}
}
\put(230,-14.5)
{
\scalebox{0.3}{\includegraphics{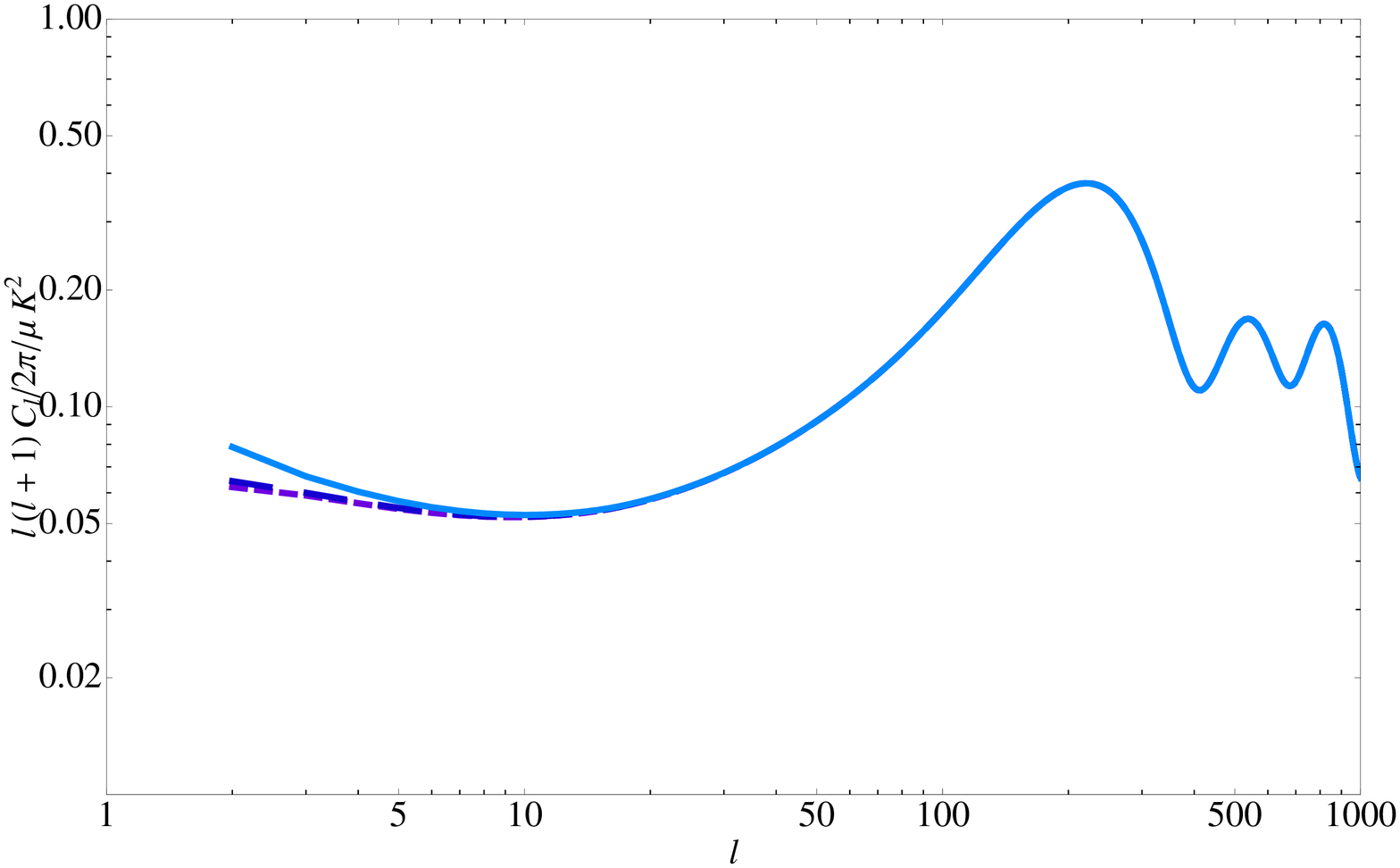}}
}
\put(90,135){\textbf{(a)}}
\put(350,135){\textbf{(b)}}
\end{picture}
\caption{Primordial $P_{\zeta}(k)$ (a) and angular $C_l$ (b) power
spectra. Here, we set the slow-roll parameters as the typical values
$\epsilon_0=\eta_0=0.01$ and $\sigma=2$. In our calculations, the
pivot wavenumber $k_0$ equals to $0.002{\rm Mpc}^{-1}$.
}\label{fig_power}
\end{figure}
\end{widetext}

Finally, we can get the primordial power spectrum of curvature
perturbations as
\begin{equation}
P_{\zeta}(k)\equiv\frac{k^3}{2\pi^2}\langle\zeta^2\rangle\approx\frac{H^2}{4\pi^2\epsilon_0}(1+C\frac{k_0}{k})
\;.
\end{equation}
This result is in agreement with that in \cite{bojowald11c}. When
all the corrections vanish, this result is back to the case of the
single field inflationary model in Einstein gravity
\cite{arXiv:0209156,maldacena03,Seery:2005wm,Chen:2006nt,Chen:2006xjb}.
The primordial power spectrum of curvature perturbations and angular
power spectrum are plotted in Fig. \ref{fig_power}, where the dotted
(purple), dashed (deep blue) and the solid (light blue) curves
correspond to different values of the parameter $C$
($C=0,4\times10^{-4},3\times10^{-3}$) which was defined after
Eq.(\ref{F}).

\section{Bispectrum}
\label{sec:bispectrum} In this section, we firstly derive the third
order Hamiltonian in the spatially flat gauge; then calculate the
three-point functions of the primordial curvature perturbations; and
finally figure out the sizes and shapes of the bispectrum.
\subsection{The third order Hamiltonian and In-In formalism}
We can get the corrected third order Hamiltonian
\begin{eqnarray}
\label{Q_ham3} \mathscr{H}^{(3)}&=&\mathscr{H}_{\rm
grav}^{(3)}[\bar{N}]+\mathscr{H}_{\rm grav}^{(3)}[\delta
N]+\mathscr{H}_{\rm matter}^{(3)}[\bar{N}]+\mathscr{H}_{\rm
matter}^{(3)}[\delta
N]\nonumber\\
&=&\frac{1}{2}\int_{\Sigma} d^3x\bar{N}\bar{\alpha}\mathfrak{H}_{\rm
grav}^{(3)}+\frac{1}{2}\int_{\Sigma} d^3x\delta N
\bar{\alpha}\mathfrak{H}_{\rm
grav}^{(2)}+\int_{\Sigma}d^3x\Big[\bar{\nu}
\mathfrak{H}_{\rm \pi}^{(3)}+\bar{\theta}\mathfrak{H}_{\rm \nabla}^{(3)}+\mathfrak{H}_{\rm \varphi}^{(3)}\Big]\nonumber\\
&+&\int_{\Sigma}d^3x\delta N[\bar{\nu}\mathfrak{H}_{\rm
\pi}^{(2)}+\mathfrak{H}_{\rm \varphi}^{(2)}+\bar{\theta}
\mathfrak{H}^{(2)}_{\nabla}]\;,
\end{eqnarray}
where the expressions for $\mathfrak{H}_{grav}^{(3)}$, \
$\mathfrak{H}_{\pi}^{(3)}$,\ $\mathfrak{H}_{\nabla}^{(3)}$,\
$\mathfrak{H}_{\varphi}^{(3)}$ are complicated, and we list them in
Appendix \ref{inter_ham}.

The perturbed third order diffeomorphism constraint is
\begin{eqnarray}
\label{Q_diff3} \mathscr{D}^{(3)}[\delta
N^a]&:=&\frac{1}{\gamma}\int_{\Sigma}d^3x\delta
N^a\Big[\frac{1}{2\bar{p}}\epsilon^{ij}_{b}(\partial_a\partial_c\delta
E^c_i)\delta E^b_j+\gamma\partial_a\delta K^j_b\delta
E^b_j-\frac{1}{2\bar{p}}\epsilon^{jk}_{a}(\partial_b\partial_c\delta
E^c_k)\delta E^b_j\nonumber\\
&-&\gamma\partial_b\delta K^j_a\delta
E^b_j-\frac{1}{2\bar{p}}\epsilon^{jk}_{a}\partial_c\delta
E^c_k\partial_b\delta E^b_j-\gamma\delta K^j_a\partial_b\delta
E^b_j+\delta\pi\partial_a\delta\varphi\Big]\;.
\end{eqnarray}

We calculate the non-Gaussianity in the interaction picture \cite{Weinberg:2005vy}
\begin{equation}
\label{inin1} \langle\zeta^3(\tau)\rangle=\langle U^{-1}_{{\rm
int}}\zeta^3(\tau)U_{{\rm int}}(\tau,\tau_0)\rangle,\ \ U_{{\rm
int}}=e^{-i\int_{\tau_0}^{\tau}\mathscr{H}_{{\rm
int}}(\tau')d\tau'}\;.
\end{equation}
Up to the first order, we have
\begin{equation}
\label{inin2}
\langle\zeta^{3}(\tau)\rangle=-i\int_{\tau_0}^{\tau}d\tau'\langle[\zeta^{3}(\tau),\mathscr{H}_{{\rm
int}}(\tau')]\rangle.
\end{equation}

\subsection{Sizes and shapes}
Based on the second order anomaly free perturbative LQC theory,
we combine Eq.(\ref{Q_ham3}) with Eq.(\ref{Q_diff3}) and arrive at
the third order interaction Hamiltonian with some counterterms
\begin{eqnarray}
\label{ham_3rd} \mathscr H^{(3)}_{{\rm int}}={\mathscr
H}^{(3)}+\mathscr D^{(3)}&=&\int
d^3x\Big[\frac{\bar{p}}{2\bar{\nu}}(1-g_1)\delta{\varphi'}^{2}\phi+\frac{\bar{p}\bar{\theta}}{2}(1+g_5)\delta^{ab}\partial_a\delta
\varphi\partial_b\delta \varphi \phi\nonumber\\
&&+\frac{\bar{p}}{\bar{\nu}}(1-g_1)(\partial^a B)\delta
{\varphi'}\partial_a \delta \varphi-
\frac{3\bar{p}}{\bar{\alpha}}(1+2f){\mathcal
H}^2\phi^3\nonumber\\
&&+\frac{\bar{p}}{2\bar{\nu}}(1+2f_1-g_1){\bar{\varphi}}^{'2}\phi^3
+\frac{(1+g_6)\bar{p}^2}{2}V_{,\rm
\varphi\varphi}(\bar{\varphi})\delta \varphi^2
\phi\nonumber\\
&&+\frac{\bar{p}^2}{6}V_{,\varphi\varphi\varphi}(\bar{\varphi})\delta\varphi^3
-\frac{(1+f_1-g_1)\bar{p}}{\bar{\nu}}{\bar{\varphi}'}\delta{
\varphi'}\phi^2+...\Big]\;,
\end{eqnarray}
where the first three terms are in the leading order under the
slow-roll approximation. In the following calculations, we only
consider these terms. Comparing Eq. (\ref{ham_3rd}) with those in
\cite{arXiv:0209156,maldacena03,Seery:2005wm,Chen:2006nt,Chen:2006xjb},
we could attribute two kinds of modifications in the interaction
Hamiltonian to the inverse volume corrections. One comes from the
modification of vacuum state (\ref{sol_vacu}), the other from the
background dependent coefficients, such as ($\bar
\nu,\bar\theta,\bar\alpha,g_1,\cdots $). However, the non-Gaussian
signatures from the latter are contaminated greatly by the cosmic
variance. Hence, we ignore the modifications in $\bar\nu$ {\it
etc.}, when we calculate the parts of three-point functions directly
from the In-In formalism. In another words, we ignore the quantum
anomaly behavior for the Hamiltonian in this work. In order to
demonstrate the reasons more manifestly, we take the first term in
(\ref{ham_3rd}) as an example.
\begin{eqnarray}
&&\int d^3x\frac{\bar{p}}{2\bar{\nu}}(1-g_1)\delta{\varphi'}^{2}\phi
\nonumber\\
&=&\int d^3x \frac{\bar{\varphi'}}{\mathcal
{H}}\frac{\bar{p}}{4}{\delta\varphi'}^2\delta\varphi +\int d^3x
\frac{\bar{\varphi'}}{\mathcal
{H}}\frac{\bar{p}}{4}\Big[\frac{5\alpha}{2}-2\nu-\nu\left(\frac{\sigma}{3}+1\right)\Big]\delta_{\text{inv}}{\delta\varphi'}^2\delta\varphi\;,
\end{eqnarray}
where the first term in the second line appears in the usual form,
while the second one contains $\delta_{\text{inv}}=\delta_{max}
\tau^{\sigma}$. Note $\delta_{\text{inv}}$ depends on $k$, there is a
corresponding $\bar{k}$ related to $\delta_{max}$. When we put these
terms into the formalism (\ref{inin2}) and perform the time
integral, we will obtain such corrections
\begin{equation}
\label{term_ign} \frac{\bar{k}^2}{K^3}\mathcal
G(k_1,k_2,k_3)+\cdots\;,\quad K\equiv k_1+k_2+k_3\;,
\end{equation}
where the dots denotes higher order corrections. From the above
expression, we can see that such term peaks at $\bar{k}\gg K$. Since
larger $k$ gives smaller $\delta_{{\rm Pl}}$, and quadruple is the
lowest detectable mode in CMB, $\bar{k}$ corresponds to $\bar{l}=2$.
So Eq.(\ref{term_ign}) peaks on the very low $l\ll2$ region where
the cosmic variance dominates over the signals (See Figure 1 and
Figure 2 in \cite{bojowald11c}). Although the non-Gaussian features
are presented on both large scale and small scale, above analysis
implies that we can ignore the effects on small scales. Hence, we
can ignore such terms in our following calculations. Once we ignore
the corrections in the background dependent coefficients in
Eq.(\ref{ham_3rd}), the form of the third order interaction
Hamiltonian reduces to the usual one
\cite{arXiv:0209156,maldacena03,Seery:2005wm,Chen:2006nt,Chen:2006xjb}.

In order to eliminate the terms proportional to the linear equations, we
need to do the field redefinition
\begin{eqnarray}
\zeta&=&\zeta_c-\frac{1}{2}\left(1-\frac{{\bar{\varphi}^{''}}}{{\bar{\varphi}^{'}}{\mathcal
H}}\right)\zeta_c^2+\frac{1}{8}\frac{{\bar{\varphi}}^{'2}}{{\mathcal
H}^2}\zeta_c^2+\frac{1}{4}\frac{{\bar{\varphi}}^{'2}}{{\mathcal
H}^2}\partial^{-2}(\zeta_c\partial^2\zeta_c)\;,\nonumber\\
&=&\zeta_c+C_1\zeta_c^2+C_2\partial^{-2}(\zeta_c\partial^2\zeta_c)\;.
\label{field_redefinition}
\end{eqnarray}
Then, the interaction Hamiltonian can be reduced into the simple
form
\begin{eqnarray}
\mathscr {H}^{(3)}_{{\rm int}}\simeq-\int d^3x
\frac{\bar{\varphi}^{'4}}{{\mathcal
H}^3}\bar{p}\zeta^{'2}_{c}\partial^{-2}\zeta^{'}_c+\cdots\;.\label{third}
\end{eqnarray}

According to
\cite{arXiv:0209156,maldacena03,Seery:2005wm,Chen:2006nt,Chen:2006xjb},
after a field redefinition of the schematic form
$\zeta=\zeta_c+\lambda \zeta_c^2$ then the correlation function will
contain two terms
\begin{eqnarray}
\langle\zeta(x_1)\zeta(x_2)\zeta(x_3)\rangle=\langle\zeta_c(x_1)\zeta_c(x_2)\zeta_c(x_3)\rangle+
2\lambda\Big[\langle\zeta(x_1)
\zeta(x_2)\rangle\langle\zeta(x_1)\zeta(x_3)\rangle+{\rm cyclic}\Big]\;,
\end{eqnarray}
where the first term can be computed by the In-In formalism and the second term
comes from the field redefinition $\zeta=\zeta_c+\lambda \zeta_c^2$.

Let us firstly calculate the first term
\begin{eqnarray}
\label{3pt}
\langle\zeta_c(\tau,k_1)\zeta_c(\tau,k_2)\zeta_c(\tau,k_3)\rangle&=&-i\int_{-\infty}^{0}
d\tau\langle0|\left[\zeta_c(\tau,k_1)\zeta_c(\tau,k_2)\zeta_c(\tau,k_3),\mathcal
{H}_{{\rm int}}^{(3)}(\tau)\right]|0\rangle\nonumber\\
&=&\frac{(2\pi)^3\delta(\sum\vec{k_i})H^6}{\left[\prod_{i=1}^{3}z^2(k_i)
2k_i^3\right]}\frac{\dot{\bar{\varphi}}^4}{H^4}\frac{4}{H^2}
\frac{\sum_{i>j}k_i^2k_j^2}{K}\left\{\prod_{j=1}^{3}\left[1+C\left(\frac{k_0}{k_j}\right)^\sigma\right]^2\right\}\;,\nonumber\\
\end{eqnarray}
where $z(k)$ and $C$ are defined in Eqs. (\ref{z}) and (\ref{F}),
respectively. Here, we emphasize again that corrections from the
background dependent coefficients are neglected in the above
expressions. However, in the next calculations of the parts from
field redefinition, such corrections should be taken into account
for consistence. Because the background dependent coefficients
(${\bar\varphi}^{'}/\mathcal H|_{\ast}\;,\cdots$)  take the value at
the moment when the corresponding mode crosses horizon
($\tau_{\ast}\sim k^{-1}$), they contain corrections such as
$(k_0/k_i)^{\sigma}$. Of course, these terms also peak at the points
where $k_0\gg k_i$, they become important, particularly, in the
squeezed triangle limit ($k_1\ll k_2,k_3$).

The contributions from field redefinition can be decomposed into two
parts, one is
\begin{eqnarray}
(2\pi)^3\delta(\sum\vec{k_i})\frac{4H^4}{\left[\prod_{i=1}^{3}z^2(k_i)
2k_i^3\right]} \left[\sum_{i=1}^{3}C_1(k_i)k_i^3
z^2(k_i)\right]\;,\label{unexpand:2}
\end{eqnarray}
and the other is
\begin{eqnarray}
(2\pi)^3\delta(\sum\vec{k_i})\frac{2H^4}{\left[\prod_{i=1}^{3}z^2(k_i)
2k_i^3\right]} \left[\sum_{i\neq j}C_2(k_i)z^2(k_i)k_i
k_j^2\right]\;,\label{unexpand:3}
\end{eqnarray}
where $H=\frac{\dot{\bar{p}}}{2\bar{p}}$ and the overdot stands for
derivative with respect to the cosmic time. The $C_1(k_i),C_2(k_j)$
terms can be read from Eq.(\ref{field_redefinition}) by substituting
$\tau$ with $k^{-1}_i$.

In summary, we conclude that the forms of interaction Hamiltonian
are exactly the same as the usual one
\cite{arXiv:0209156,maldacena03,Seery:2005wm,Chen:2006nt,Chen:2006xjb},
however, the inverse volume corrections $\delta_{{\rm inv}}$ will
make some contribution to the bispectrum. There are mainly two
sources, one is the modifications to the standard Bunch-Davies
vacuum, the other comes from the $z(k)$ factor in the gauge
transformation $\zeta(k)=u(k)/z(k)$. Furthermore, we can expand the
above results in terms of $\delta_{{\rm
inv}}=\delta_0(k_0/k_i)^{\sigma}$ and obtain
\begin{eqnarray}
(2\pi)^3\delta(\sum
\vec{k_i})\Big[\mathcal F_{{\rm single}}(k_1,k_2,k_3)
+\mathcal F_1(k_1,k_2,k_3)+\mathcal F_2(k_1,k_2,k_3)\Big]\;,\label{full_shape}
\end{eqnarray}
where
\begin{eqnarray}
\label{single_shape}
{\mathcal
F_{{\rm single}}}=\frac{(2\pi)^4P^2_{\zeta}}{4\Big[\prod_{j=1}^3(2k_j^3)\Big]}\left\{(3\epsilon_0-2\eta_0)\sum_ik_i^3+
\epsilon_0\sum_{i\neq
j}k_ik_j^2+8\epsilon_0\frac{\sum_{i>j}k_i^2k_j^2}{K}\right\}\;,\nonumber\\
\end{eqnarray}
is the usual leading term and
\begin{eqnarray}
\label{F1_shape}
\mathcal F_1&=&\left[\omega_1\sum_{l=1}^3
\left(\frac{k_0}{k_l}\right)^{\sigma}\right]\mathcal F_{{\rm single}}\;,\\
\label{F2_shape}
{\mathcal
F}_2&=&\frac{(2\pi)^4P_{\zeta}^2}{4\Big[\prod_{j=1}^3(2k_j^3)\Big]}\left\{\Big[2\omega_3'(\epsilon_0-\eta_0)+\omega_2'\epsilon_0\Big]
\sum_ik_i^3\left(\frac{k_0}{k_i}\right)^\sigma
+\omega_2'\epsilon_0\sum_{i\neq
j}k_j^2k_i\left(\frac{k_0}{k_i}\right)^{\sigma}\right\}\;,\nonumber\\
\end{eqnarray}
are the inverse volume correction terms, and here the relevant
coefficients are
\begin{eqnarray}
\omega_1&=&2C-C_z\;,\\
\omega_2^{\prime}&=&\gamma_{\epsilon}+C_z-2C\;,\\
\omega_3^{\prime}&=&\gamma_{\eta}+C_z-2C\;,\\
C_z&=&-\delta_0\left[\nu_0\left(\frac{\sigma}{6}+1\right)+\frac{\sigma\alpha_0}{2\epsilon_0}\right]\;,\\
C&=&-\frac{\delta_0\chi}{2(\sigma+1)}\;.
\end{eqnarray}

Particularly, we figure out that the corrections from
$\Big[\prod_{i=1}^3z^2(k_i)\Big]$ terms in the denominators in
Eqs.(\ref{3pt}), (\ref{unexpand:2}) and (\ref{unexpand:3}) are
absorbed into $\mathcal F_1$ shape, while all other corrections are
collected in $\mathcal F_2$. In the next section, we will find
$\mathcal F_1$ shape provides an unique signal from LQC mechanism,
and more importantly, this signal is independent of the inflationary
models, because $\Big[\prod_{i=1}^3z^2(k_i)\Big]$ terms always
appear in the gauge transformations (\ref{z}). Namely it is an
universal signal in the LQC scenario.

The single shape is the usual one, while $\mathcal F_1$ and
$\mathcal F_2$ arise only in LQC scenarios. Sizes of the two new
parts are proportional to parameters ($\omega_1\;,\cdots$), which
are in the order of $\mathcal O(10^{-3},10^{-4})$. That is to say
that, these new non-Gaussian features from LQC are smaller than
those of usual inflationary models in Einstein gravity by a factor
at least $10^{-3}$. Although this does be a tiny number, considering
these features sourced by quantum effect, this factor is not small
as initially expected. Furthermore, as stated above, because the
inverse volume corrections in the interaction Hamiltonian can be
neglected, i.e. we could use the usual Hamiltonian directly in the
In-In formalism by substituting the Bunch-Davies vacuum state with
the one given in Eq.(\ref{sol_vacu}), we argue that the expectations
on the sizes of bispectrum should hold for any inflationary models
in LQC scenario.

\begin{widetext}
\begin{figure}
\begin{picture}(500,190)(0,0)
\put(-15,-15)
{
\scalebox{0.4}{\includegraphics{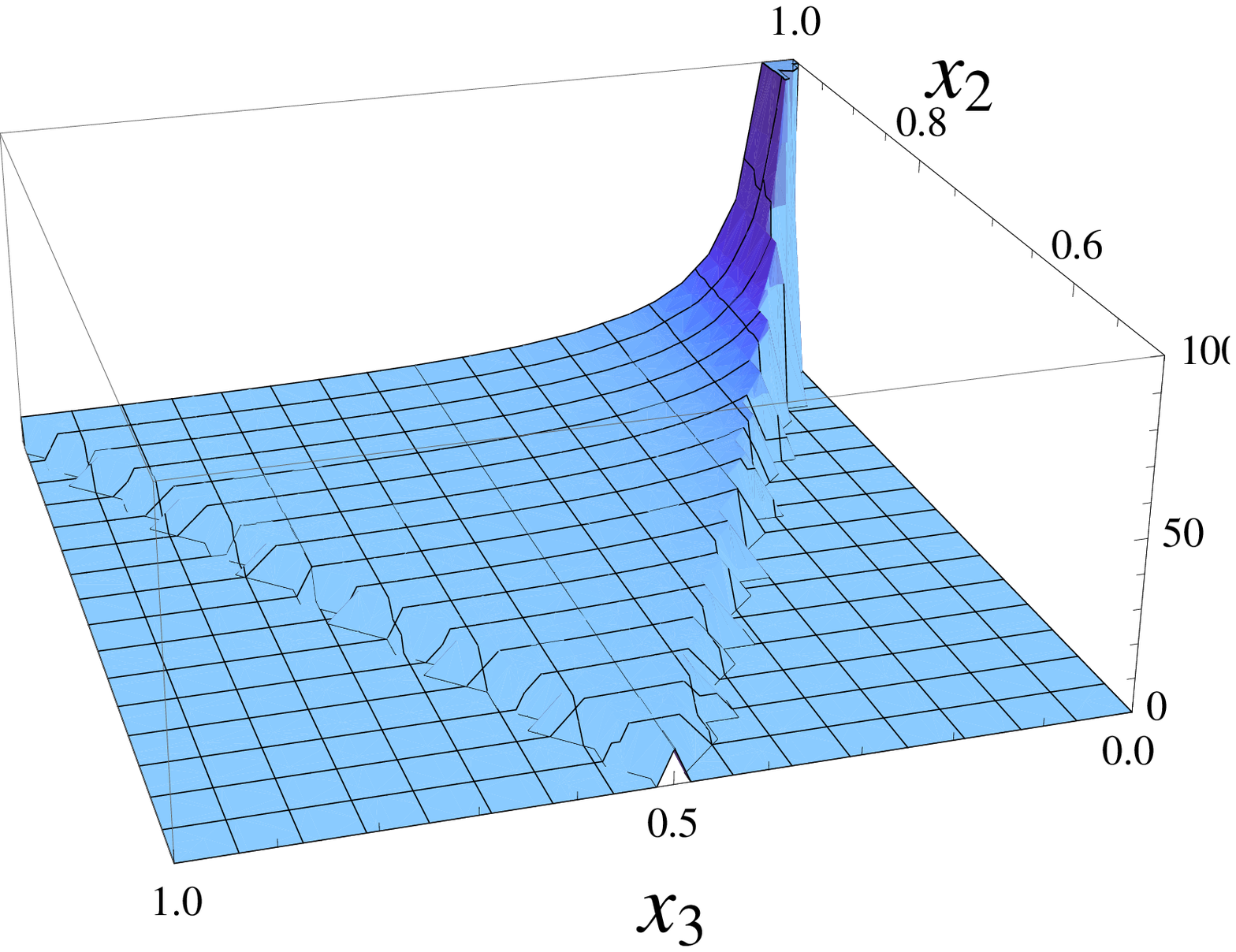}}
}

\put(230,-14.5)
{
\scalebox{0.4}{\includegraphics{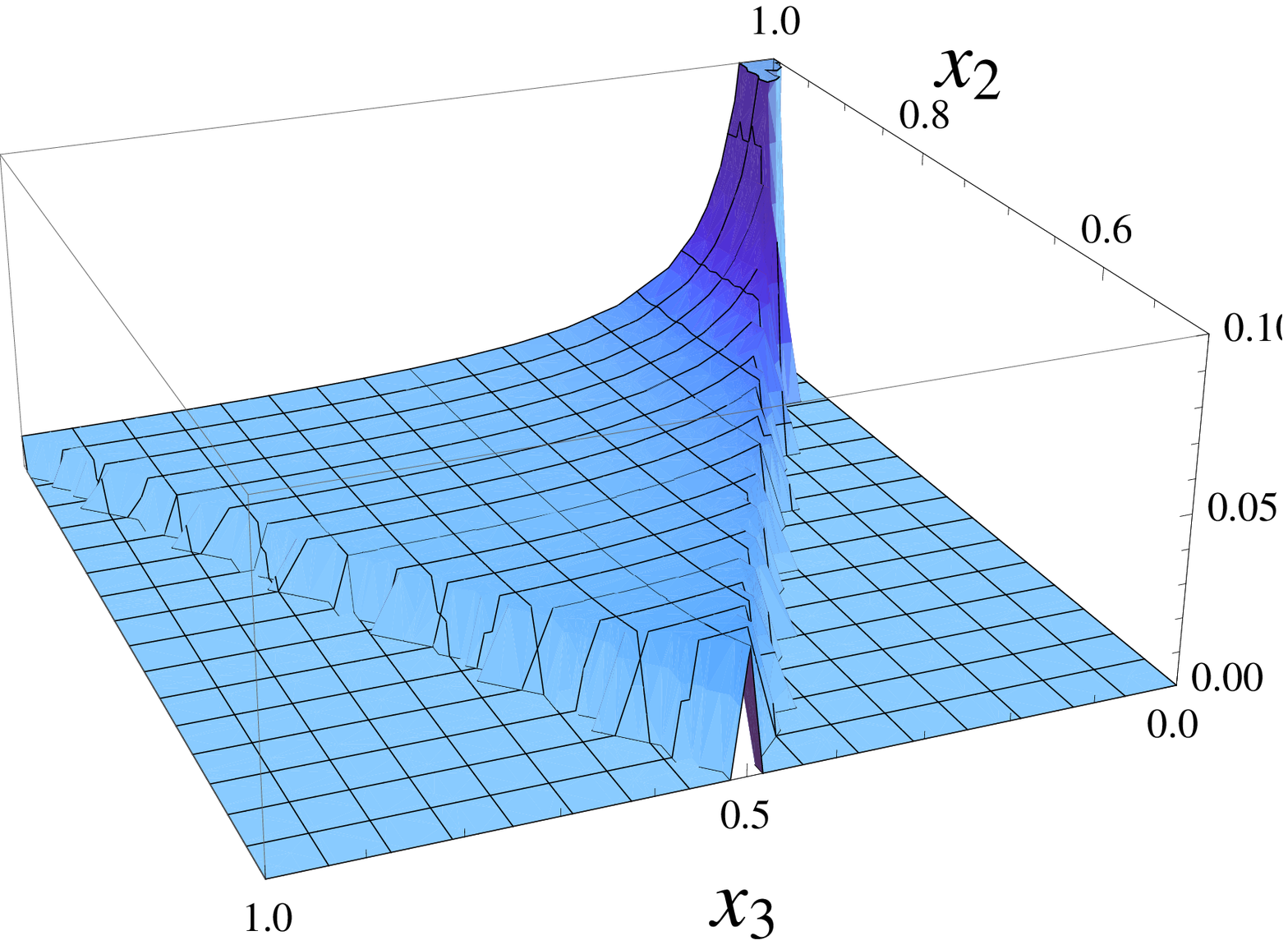}}
}
\put(45,135){\textbf{(a)}}
\put(280,135){\textbf{(b)}}
\end{picture}
\caption{Single $({\rm a})$ and $\mathcal F_1$ $({\rm b})$ shapes.
The $z$-axis is $x_2^2x_3^2\mathcal F(x_2,x_3)/P_{\zeta}^2$, here we
set $x_1=1$. The slow-roll parameters take the typical values
$\epsilon_0=\eta_0=0.01$. Here and hereafter we set $\sigma=2$.
}\label{fig_bisp1}
\end{figure}
\end{widetext}

\begin{widetext}
\begin{figure}
\begin{picture}(500,190)(0,0)
\put(-15,-15)
{
\scalebox{0.4}{\includegraphics{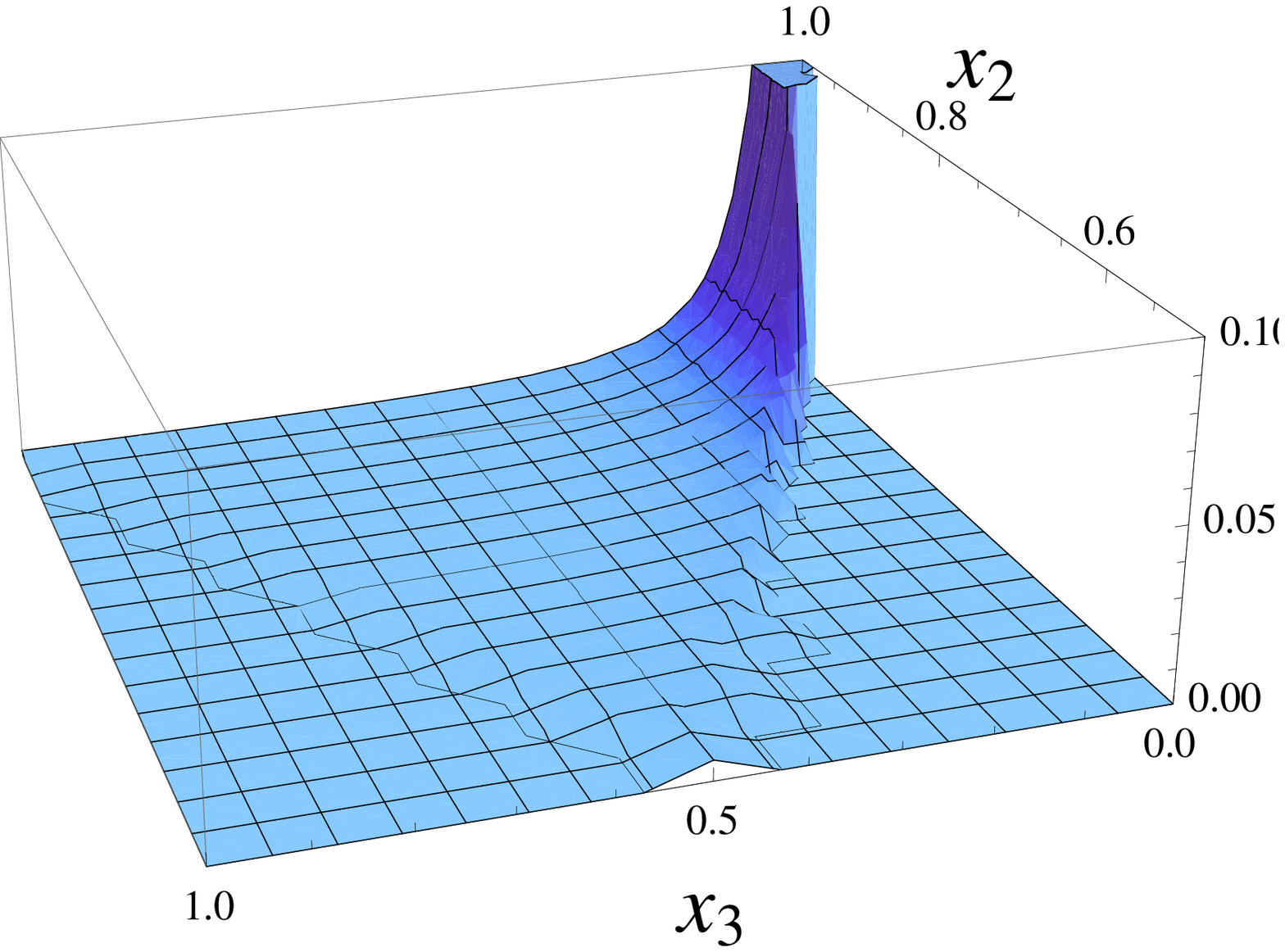}}
}

\put(230,-14.5)
{
\scalebox{0.4}{\includegraphics{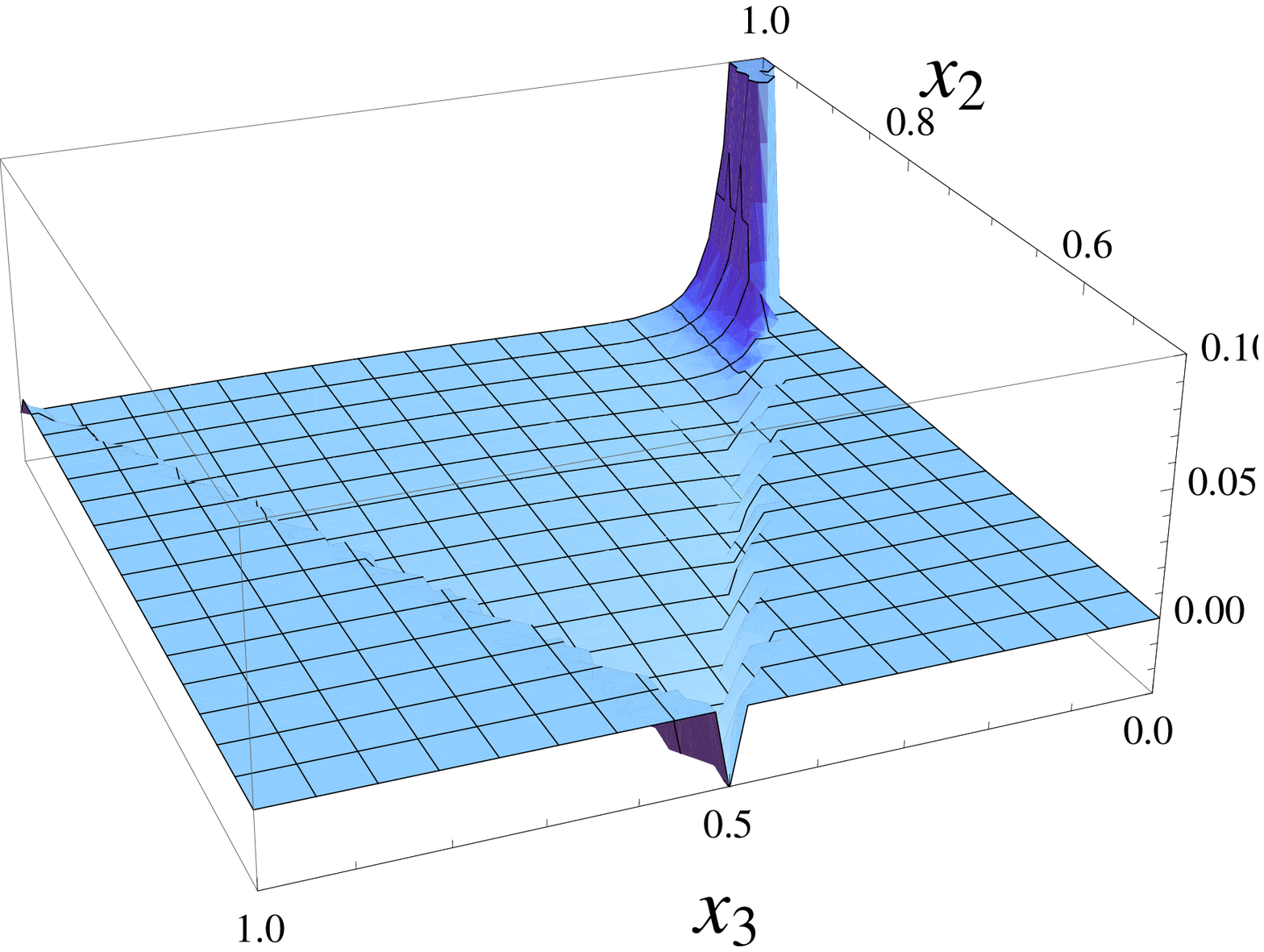}}
}
\put(45,135){\textbf{(c)}}
\put(280,135){\textbf{(d)}}
\end{picture}
\caption{$\mathcal F_2$ (c) shape and $\mathcal F_2-\mathcal F_1$ (d).
The coefficients ($\omega_1\;,\omega_2'\;,\omega_3'$) in $\mathcal F_1,\mathcal F_2$ are set unity.}\label{fig_bisp2}
\end{figure}
\end{widetext}

In Fig. \ref{fig_bisp1} (a) (b) and Fig. \ref{fig_bisp2} (c), we
plot $x_1^2x_2^2x_3^2\mathcal F(x_1,x_2,x_3)/P^2_{\zeta}$, with
$x_i\equiv k_i/k_1$. The difference between shapes $\mathcal F_1$
and $\mathcal F_2$ is plotted in Fig. \ref{fig_bisp2} (d). We can
see that all the three shapes peak at the squeezed limit
($x_2=1,x_3=0$), however, the substructures are different among
them. Compared to the single shape, $\mathcal F_1$ shape upraises at
the corner ($x_2=0,x_3=0.5$), while $\mathcal F_2$ flattens at the
same point. From Fig. \ref{fig_bisp2} (d), we can also see that
$\mathcal F_2$ peaks more dramatically in the squeezed corner than
$\mathcal F_1$.

\section{Shape correlations}
\label{sec:shape}
In the Figures \ref{fig_bisp1} and \ref{fig_bisp2}, all the three shapes looks
similar. In order to figure out the differences among shapes more quantitatively,
we need to calculate the correlations between them.

Firstly, we define a 3D shape function \cite{arXiv:0812.3413,astro-ph/0405356,arXiv:1006.2771}
\begin{equation}
\label{shape_func}
S(k_1,k_2,k_3)=\frac{1}{N}(k_1k_2k_3)^2\mathcal F(k_1,k_2,k_3)\;,
\end{equation}
where $N$ is a normalization factor which will not affect the
following calculations.

Then, we construct the products of two shape functions
\begin{equation}
\label{shape_prod}
F(S,S')=\int_{\mathcal V_k}S(k_1,k_2,k_3)S'(k_1,k_2,k_3)\omega(k_1,k_2,k_3)d\mathcal V_k\;,
\end{equation}
where $\omega(k_1,k_2,k_3)$ is a weight function and $\mathcal V_k$
is the integration domain constrained by the triangle inequality.

Finally, we arrive at the 3D shape function correlator
\begin{equation}
\label{shape_corr}
\bar{\mathcal C}(S,S')=\frac{F(S,S')}{\sqrt{F(S,S)F(S',S')}}\;.
\end{equation}
This quantity describe the cross correlations between two different shapes.

Furthermore, for a large series of well-motivated shapes, the above
descriptions can be simplified. One can define the distance
described by $k$ from the origin of ($k_1,k_2,k_3$)-momentum space
to the particular triangle slice which is perpendicular to ($1,1,1$)
direction,
\begin{equation}
k\equiv\frac{1}{2}(k_1+k_2+k_3)\;,
\end{equation}
then, we introduce another two new variables
\begin{eqnarray}
\label{alpha_beta}
k_1&=&k(1-\beta)\;,\\
k_2&=&\frac{1}{2}k(1+\alpha+\beta)\;,\\
k_3&=&\frac{1}{2}k(1-\alpha+\beta)\;.
\end{eqnarray}
In the domain constrained by the triangle inequality, $0\leq
k\leq\infty$, $0\leq\beta\leq1$ and
$-(1-\beta)\leq\alpha\leq1-\beta$. For the classes of models with
homogeneous shape, which means that the powers of wavenumber in
shapes are homogeneous, the $k$ dependence in the 3D shape function
can be separated
\begin{equation}
S(k_1,k_2,k_3)=f(k)\mathscr S(\alpha,\beta)\;,\qquad d\mathcal V_k=dk_1dk_2dk_3=k^2dkd\alpha d\beta\;.
\end{equation}
In fact, for the models considered in
\cite{arXiv:0812.3413,astro-ph/0405356,arXiv:1006.2771} and here,
$f(k)\propto {\rm const.}$ Hence, we can focus on the 2D shape
function $\mathscr S(\alpha,\beta)$ and the integral over $k$
cancels when one calculates the shape function correlators.

Further, one can introduce
\begin{equation}
\alpha=1-x\;,\qquad \beta=xy/3\;,\quad (0\leq x,y\leq1)\;,
\end{equation}
to square the integration regime. By using these variables, the integral measurement
becomes
\begin{equation}
d\alpha d\beta=xdxdy\;.
\end{equation}
From the above expression, we can read the weight function
$w(x,y)=x$. This choice works well for all the shapes mentioned in
\cite{arXiv:0812.3413,astro-ph/0405356,arXiv:1006.2771}, however,
for our new shapes $\mathcal F_1$ and $\mathcal F_2$ the correlation
matrices $c_{mn}$ defined in (\ref{c_mn}), suffers from divergence.
In our calculation we therefore use a new variable $\xi^2=x$
\begin{equation}
\label{new_var}
d\alpha d\beta=\xi^3d\xi dy\;,
\end{equation}
to eliminate such divergence. Using variables ($\xi,y$), we can decompose the shape $\mathscr S(\xi,y)$
on any triangle slice with analogous radial polynomials $R_m(\xi)$ and shifted Legendre polynomials
$\bar{P}_n(y)$
\begin{equation}
\label{shape_decomp}
\mathscr S(\xi,y)=\sum_{m,n}c_{mn}R_m(\xi)\bar{P}_n(y)\;,
\end{equation}
where the first few $R_m(\xi)$ eigenfunctions are
\begin{eqnarray}
\label{eigen1}
&&R_0=\sqrt{2}\;,\quad R_1=\sqrt{4}(-2+3\xi)\;,\quad R_2=\sqrt{6}(3-12\xi+10\xi^2)\;,\quad\nonumber\\
&&R_3=\sqrt{8}(-4+30\xi-60\xi^2+35\xi^3)\;,\cdots.
\end{eqnarray}
And $\bar P_n(y)$ eigenfunctions are
\begin{eqnarray}
\label{eigen2}
&&\bar P_0=1\;,\quad \bar P_1=\sqrt{3}(-1+2y)\;,\quad \bar P_2=\sqrt{5}(1-6y+6y^2)\;,\nonumber\\
&&\bar P_3=\sqrt{7}(-1+12y-30y^2+20y^3)\;,\cdots.
\end{eqnarray}

Thus, one can define the correlation matrix $c_{mn}$
\begin{equation}
\label{c_mn}
c_{mn}=\int_0^1d\xi\int_0^1dy\xi^3\mathscr S(\xi,y)R_m(\xi)\bar P_n(y)\;.
\end{equation}
The $c_{mn}$ matrices for local, equilateral, single field model,
$\mathcal F_1$ and $\mathcal F_2$ are listed in (\ref{list:c_mn})
and Fig. \ref{fig_corr}. The explicit definitions of 2D shape
function $\mathscr S(\xi,y)$ can be found in Appendix
\ref{shapefunc_definition}.
\begin{eqnarray}
&&\left(
\begin{array}{cccc}
 1.00 & -0.14 & 0.03 & 0.00 \\
 0.38 & -0.07 & 0.02 & 0.00 \\
 0.04 & -0.01 & 0.01 & 0.00 \\
 0.02 & 0.00  & 0.00 & 0.00
\end{array}
\right)\;,
\quad
\left(
\begin{array}{cccc}
 1.00 & 0.40 & -0.12 & 0.01 \\
 0.68 & 0.27 & -0.09 & 0.01 \\
 0.21 & 0.07 & -0.03 & 0.00 \\
 0.01 & 0.00 & -0.01 & 0.00
\end{array}
\right)\;,
\quad
\left(
\begin{array}{cccc}
 1.00 & -0.07 & 0.01 & 0.00 \\
 0.43 & -0.02 & 0.01 & 0.00 \\
 0.07 & 0.00 & 0.00 & 0.00 \\
 0.02 & 0.00  & 0.00 & 0.00
\end{array}
\right)\;,\nonumber\\
&&\qquad\qquad
\left(
\begin{array}{cccc}
 1.00 & -0.19 & 0.04 & 0.00 \\
 -0.26 & 0.09 & -0.01 & 0.00 \\
 0.44 & -0.10 & 0.02 & 0.00 \\
 -0.32 & 0.08  & -0.01 & 0.00
\end{array}
\right)\;,\qquad
\left(
\begin{array}{cccc}
 1.00 & -0.12 & 0.24 & 0.00 \\
 0.30 & -0.03 & 0.01 & 0.00 \\
 0.07 & -0.01 & 0.01 & 0.00 \\
 0.02 & 0.00  & 0.00 & 0.00
\end{array}
\right)\;.
\label{list:c_mn}
\end{eqnarray}

\begin{widetext}
\begin{figure}
\begin{picture}(500,300)(0,0)
\put(0,150)
{
\scalebox{0.25}{\includegraphics{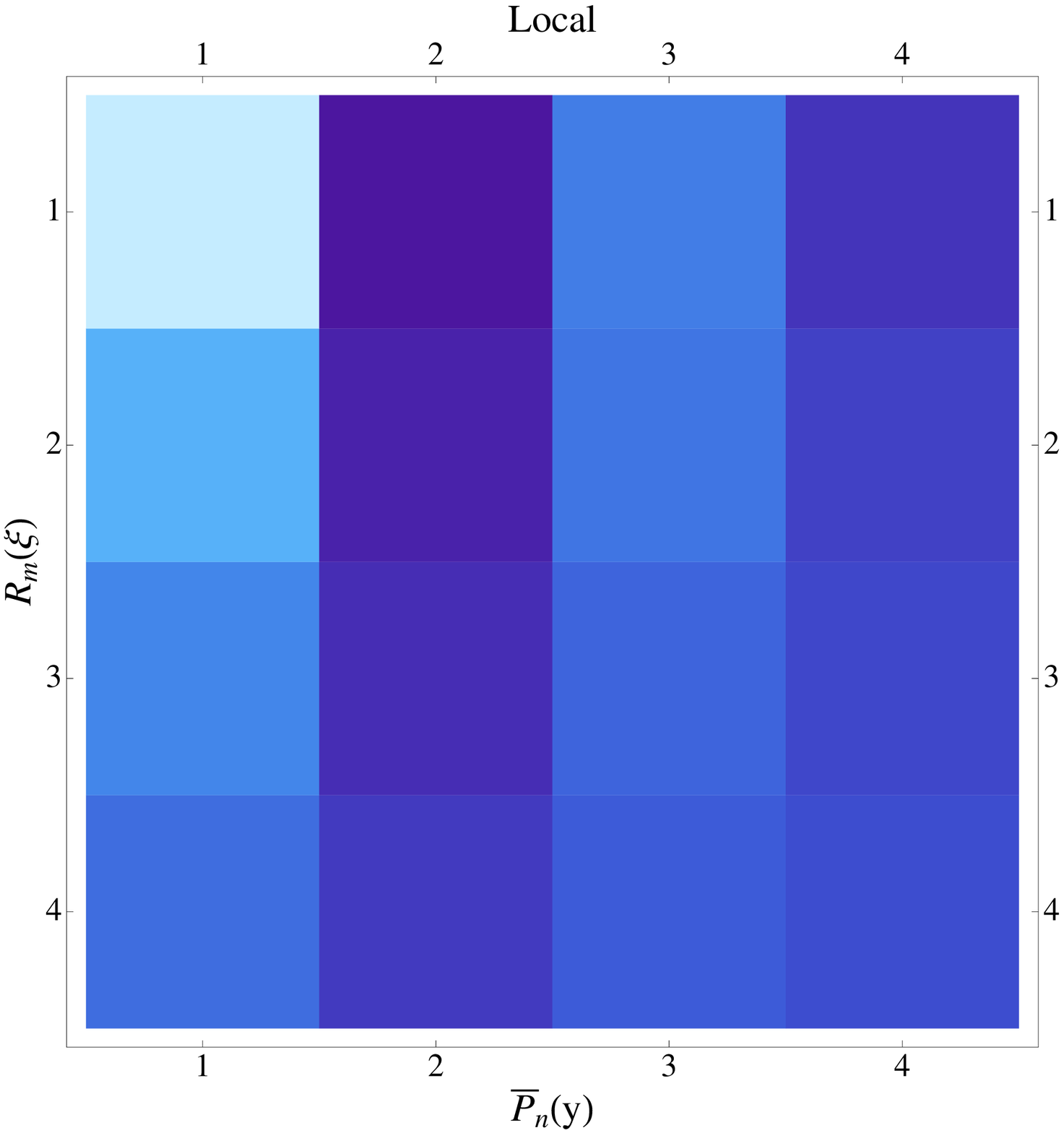}}
}

\put(150,150)
{
\scalebox{0.25}{\includegraphics{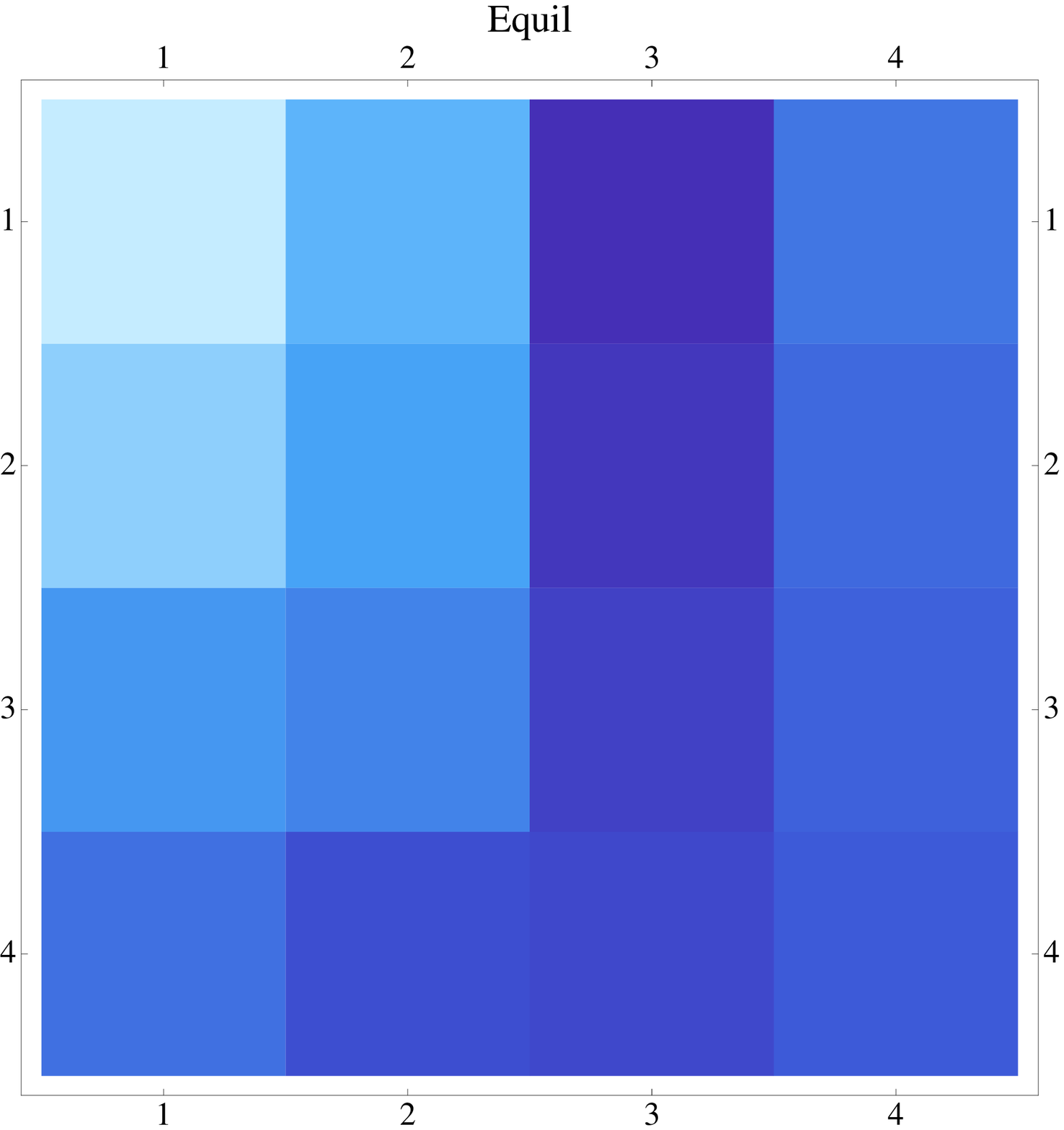}}
}

\put(300,150)
{
\scalebox{0.25}{\includegraphics{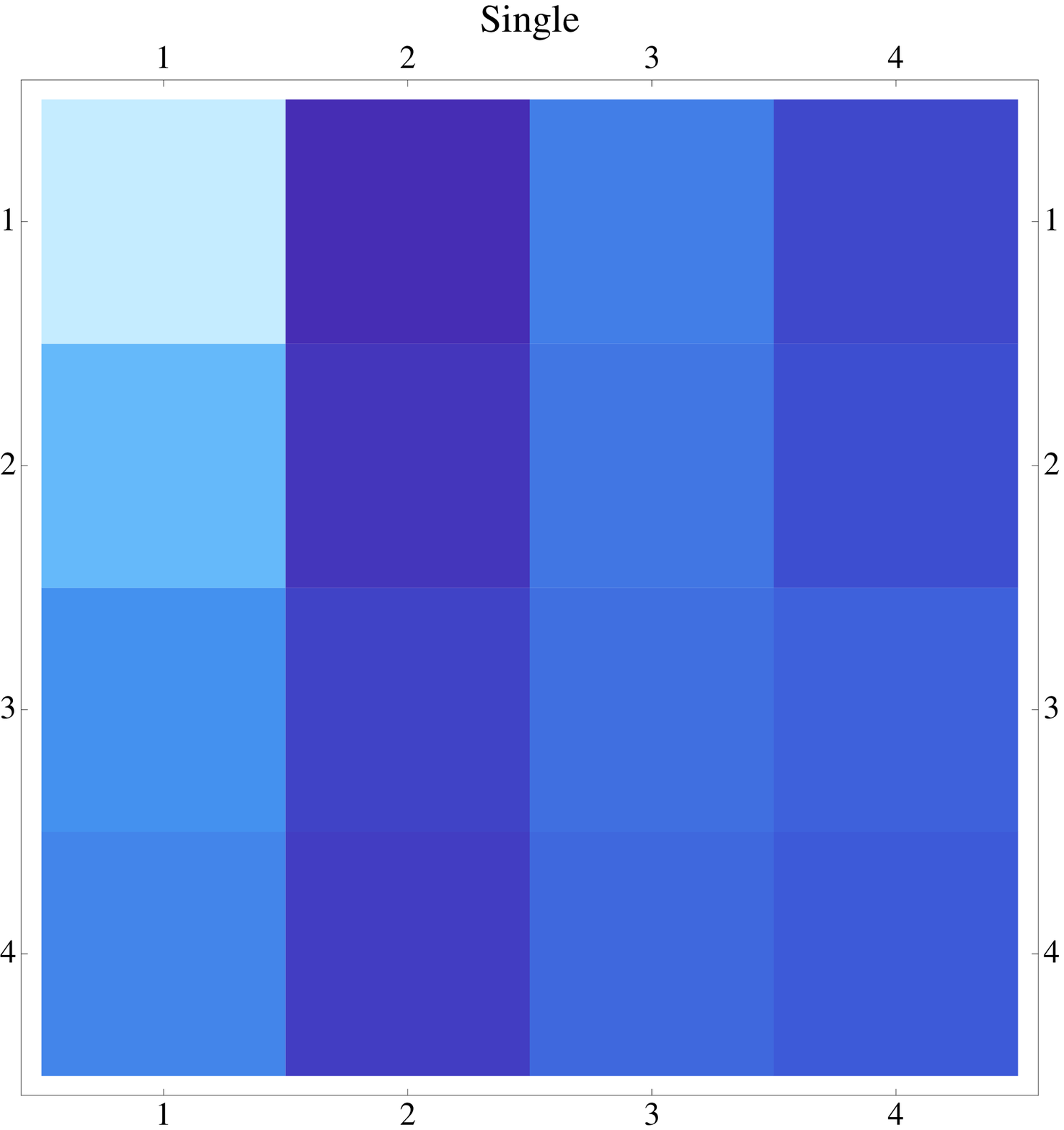}}
}

\put(50,0)
{
\scalebox{0.25}{\includegraphics{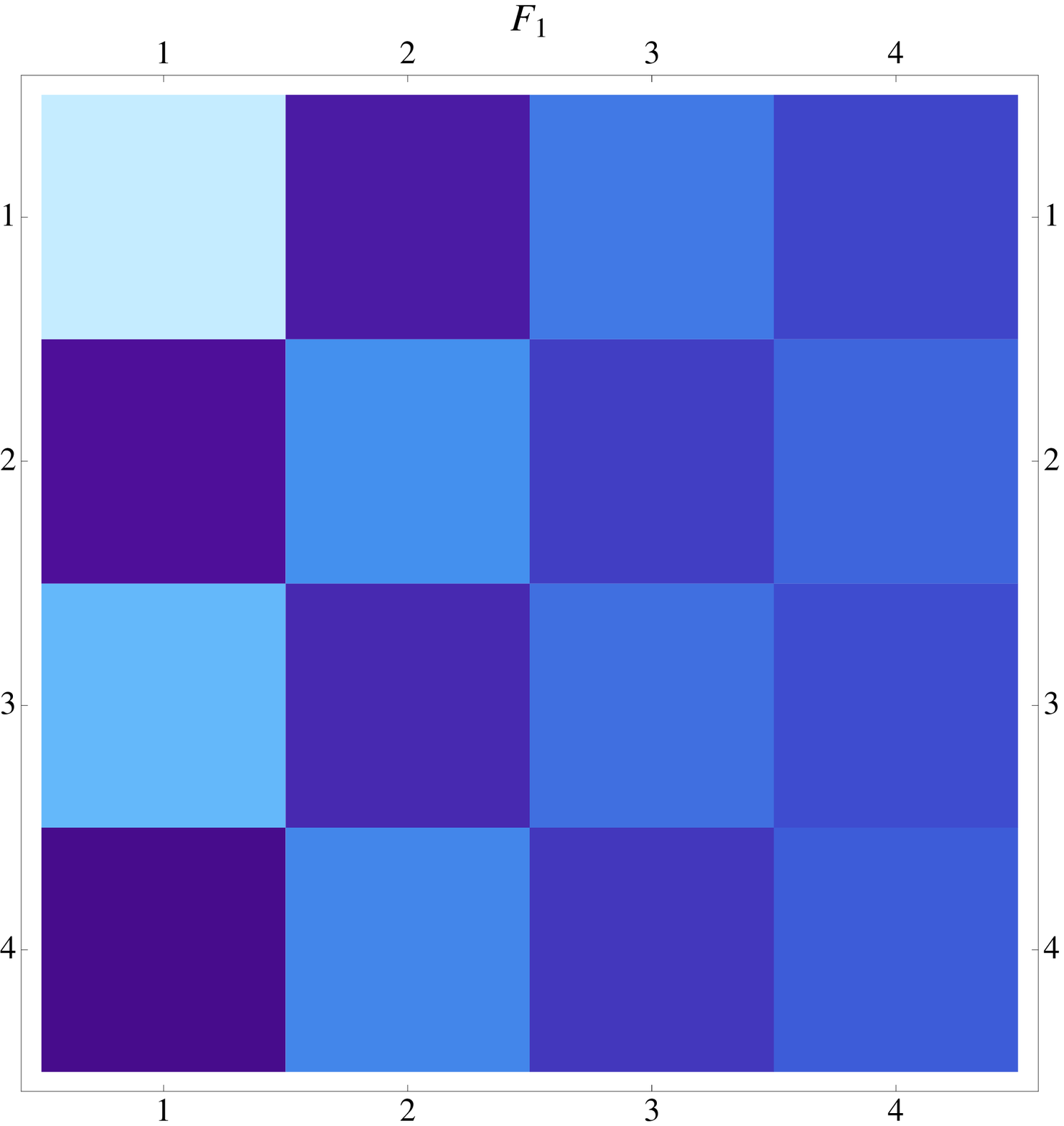}}
}

\put(250,0)
{
\scalebox{0.25}{\includegraphics{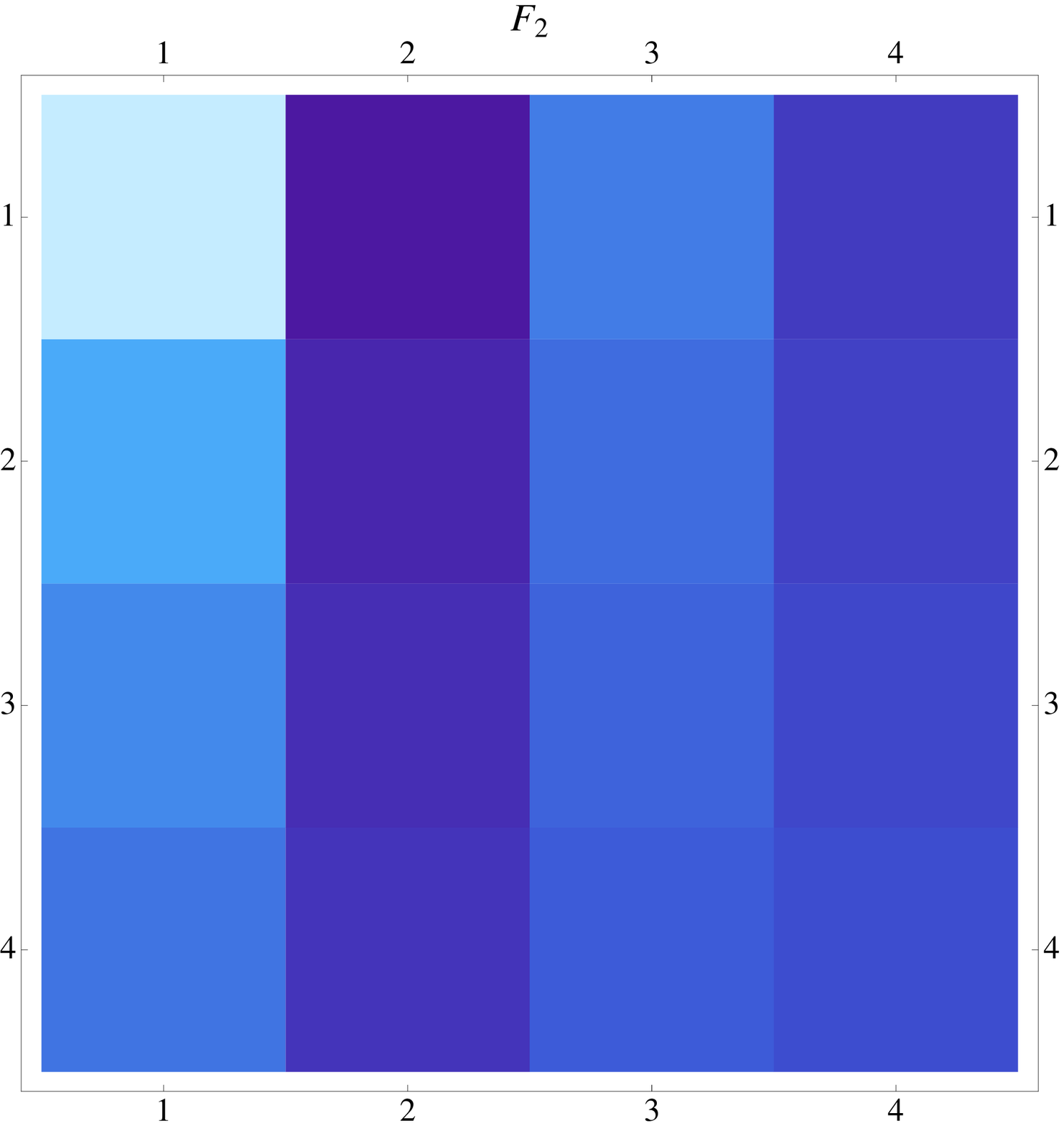}}
}
\end{picture}
\caption{Correlation matrices in (\ref{list:c_mn}). Light blue denotes
for +1, and the deep for -1.}\label{fig_corr}
\end{figure}
\end{widetext}
From Fig. \ref{fig_corr}, we can see that $\mathcal F_1$ term
differs from others explicitly, while $\mathcal F_2$ is almost
indistinguishable with the local form visually.

Armed with the above results, one can calculate 2D shape correlator
\begin{equation}
\label{corr_fact}
\mathscr C(\mathscr S,\mathscr S')\equiv\frac{\mathscr F(\mathscr S,\mathscr S')}
{\sqrt{\mathscr F(\mathscr S,\mathscr S)\mathscr F(\mathscr S',\mathscr S')}}\;,
\end{equation}
where the product is defined through
\begin{equation}
\label{corr_prod}
\mathscr F(\mathscr S,\mathscr S')\equiv\int_{\mathcal S_k}\mathscr S(\xi,y)\mathscr S'(\xi,y)
\xi^3d\xi dy=\sum_{m,n}c_{mn}c'_{mn}\;.
\end{equation}
The numerical results of the 2D correlators are listed in Tablet
\ref{tb:shapecorrelator}, from which we can find that the
correlations of shape $\mathcal F_1$ are low with all other four
shapes, while shape $\mathcal F_2$ possesses high correlations with
local (particularly), equilateral as well as single form. That is to
say that $\mathcal F_2$ is almost indistinguishable with local form,
while $\mathcal F_1$ does be the unique signal of LQC. As already
argued in the previous section, $\mathcal F_1$ is an universal shape
in LQC scenario, so we can identify this shape as a new window for
LQC scenario.

\begin{table}[t]
\centering
\begin{tabular}[t]{|@{\hspace{1mm}}c@{\hspace{1mm}}|@{\hspace{1mm}}c@{\hspace{1mm}}|@{\hspace{1mm}}c@{\hspace{1mm}}|@{\hspace{1mm}}c@{\hspace{1mm}}|@{\hspace{1mm}}c@{\hspace{1mm}}|@{\hspace{1mm}}c@{\hspace{1mm}}|}
\hline
 & Local & Equi & Single & $\mathcal F_1$ & $\mathcal F_2$ \\
\hline
Local &  1& 0.826 & 0.996 & 0.722 & 0.996 \\
Equil &  &  1& 0.874 & 0.535 & 0.826 \\
Single &  &  &  1& 0.707 & 0.992 \\
$\mathcal F_1$ &  &  &  &  1& 0.769 \\
$\mathcal F_2$ &  &  &  &  &1  \\
\hline
\end{tabular}
\caption{2D shape correlators.}
\label{tb:shapecorrelator}
\end{table}

Finally, let us estimate the parameter $f_{{\rm NL}}$. Here we focus on the two new
shapes $\mathcal F_{1,2}$, from Tab. \ref{tb:shapecorrelator} we can see that $\mathcal F_2$
is highly correlated with the local form, while $\mathcal F_1$ is less correlated with them.
The contributions to $f_{{\rm NL}}^{{\rm local}}$ from these two shapes can be easily estimated
as
\begin{equation}
\label{fnl}
\Delta f_{{\rm NL}}^{{\rm local}}\sim
\omega\times\mathscr C(\mathscr S_{{\rm local}},\mathscr S_{\mathcal F_{1,2}})
\times
f_{{\rm NL}}^{{\rm local}}({\rm single})
\sim \mathcal O(10^{-3})f_{{\rm NL}}^{{\rm local}}({\rm single})\;,
\end{equation}
where $\omega$ represents in short parameters
($\omega_1,\omega_2',\omega_3'$) in (\ref{F1_shape}) and
(\ref{F2_shape}), whose typical values are of order $\mathcal
O(10^{-3})$. Because $f_{{\rm NL}}^{{\rm local}}({\rm single})$ in
the usual case is of order $\sim \epsilon_0$~\cite{Kundu2011}, the
contributions from inverse volume corrections is completely
negligible. However, as argued before, our results should be robust
for other inflationary models in LQC scenarios, especially those
with large non-Gaussianities, such as K-inflation, DBI inflation
{\it etc.} So, one can anticipate that in those models with large
non-Gaussianities the features from inverse volume corrections in
LQC scenarios might be observed.

\section{Conclusions}
\label{sec:discussion} In this paper we investigated the
contributions to the cosmic primordial scalar bispectrum from the
inverse volume corrections in LQC scenarios. We derived the
interaction Hamiltonian, however, we found that the new interactions
contribute greatly to the modes with $k_1+k_2+k_3\ll \bar{k}$.
Because the scales corresponding to $\bar{k}$ is very large, here we
take $\bar{k}\approx0.00014{\rm Mpc}^{-1}$ corresponding to
quadruple mode $\bar{l}=2$, it means that the three wave lengthes in
the bispectrum are all on the super-horizon scales. On so large
scales the cosmic variance usually dominates over the signals, we
hence neglected these new interactions in our calculations. That is
to say, the interaction Hamiltonian we used is the same form as the
usual one for the single field inflation models in Einstein gravity.
This greatly simplifies our calculations, and more importantly,
makes our results robust, i.e., our results should hold for other
inflationary models in LQC scenarios.

Although the Hamiltonian shares the same forms as the one in
Einstein gravity, the inverse volume corrections $\delta_{{\rm
inv}}\propto(k_0/k)^{\sigma}$ still contribute to the bispectrum.
Roughly speaking, there are two aspects, one comes from the
deviations from the standard Bunch-Davies vacuum; the other
attributes to the non-trivial gauge transformations
$\zeta(k)=u(k)/z(k)$ from the spatially flat gauge to the observable
curvature perturbations. Consequently, we obtained the three-point
functions of the gauge invariant curvature perturbations. We found
that, except for the usual single component in slow-roll inflation
models, two new shapes arise due to the corrections, namely
$\mathcal F_{1,2}$. Furthermore, we performed a careful analysis on
the new shapes. We found that, the whole profiles for all the three
shapes (single, $\mathcal F_{1,2}$) are visually similar, i.e. they
peak at the squeezed limit. However, the substructures among them
are different. Compared to the single shape, $\mathcal F_1$ shape
upraises at another corner (See. Fig. \ref{fig_bisp1} and
\ref{fig_bisp2}), while $\mathcal F_2$ flattens at the same point,
and $\mathcal F_2$ peaks more dramatically in the squeezed corner
than $\mathcal F_1$. In addition, we investigated the correlations
among five shapes, including local, equilateral, single and
$\mathcal F_{1,2}$. The results show that $\mathcal F_{2}$ is highly
correlated with the local type, while $\mathcal F_{1}$ is less. It
means that the latter can provide a new window for probing the loop
quantum mechanisms using cosmic primordial bispectrum information.
Finally, we estimated the order of observable parameter $\Delta
f_{{\rm NL}}^{{\rm local}}\sim \mathcal O(10^{-3})\times f_{{\rm
NL}}^{{\rm local}}({\rm inflation})$ from the inverse volume
corrections.

The non-Gaussianity from the inverse volume corrections in LQC
scenarios is tiny and still undetectable currently, however,
considering they are generated by the quantum effect, which is
naively expected of the order $\mathcal O({\rm GUT}/{\rm
Planck})^{\sigma}\sim \mathcal O(10^{-5})^{\sigma}$, our finding
becomes non-trivial. Especially, the results obtained in this work
should be generalized directly to other inflationary models with
large non-Gaussianities, in which the inverse volume corrections
also becomes large therein. In addition, in this paper we only
investigated the non-Gaussianities from the inverse volume
corrections, while ignored those from the holonomy corrections, in
which the sound speed of scalar perturbations are typically changed.
Besides, we only investigated in this work the bispectrum from
scalar modes, left those from tensor mode unexplored. These topics
are worth investigating further.

\begin{acknowledgments}
We thank Qing-Guo Huang and Yun-Song Piao for useful discussions. BH
thanks the hospitality of ITP-CAS/KITPC during his visit. This work
is partially supported by the projects of Knowledge Innovation
Program of Chinese Academy of Science,  National Basic Research
Program of China under Grant No. 2010CB832805 and No. 2010CB833004,
and the National Natural Science Foundation of China (No. 10821504,
No. 10975168, No.11175225 and No.11035008).
\end{acknowledgments}

\appendix

\section{Interacting Hamiltonian}
\label{inter_ham}
In this appendix, we derive the perturbed Hamiltonian density and
the perturbed diffeomorphism constraint up to the third order.
According to \cite{bojowald08}, the classical Hamiltonian includes
two parts
\begin{eqnarray}
\mathscr{H}[N]=\mathscr{H}_{\rm grav}[N]+\mathscr{H}_{\rm matter}[N]
\end{eqnarray}
The gravitational Hamiltonian can be expressed in terms of the
extrinsic curvature
\begin{eqnarray}
\mathscr{H}_{\rm grav}[N]=\frac{1}{2}\int_{\Sigma}d^3x
N\mathfrak{H}=\frac{1}{2}\int_{\Sigma}d^3x N
\epsilon_i^{jk}\frac{E^c_jE^d_k}{\sqrt{|\det
E|}}\Big[2\partial_c\Gamma^i_d+\epsilon^i_{mn}(\Gamma^m_c\Gamma^n_d-K^m_cK^n_d)\Big],
\end{eqnarray}
and the matter part of the Hamiltonian is
\begin{eqnarray}
\mathscr{H}_{\rm matter}[N]=\int_{\Sigma}d^3xN(
\mathfrak{H}_{\rm\pi}+\mathfrak{H}_{\rm
\nabla}+\mathfrak{H}_{\rm\varphi}),
\end{eqnarray}
where
\begin{eqnarray}
\mathfrak{H}_{\rm\pi}=\frac{\pi^2}{2\sqrt{|\det E|}},\, \,
\mathfrak{H}_{\rm\nabla}=\frac{E^a_iE^b_i\partial_a\varphi\partial_b\varphi}{2\sqrt{|\det
E|}}, \, \, \mathfrak{H}_{\varphi}=\sqrt{|\det E|}V(\varphi).
\end{eqnarray}

The diffeomorphism constraint is
\begin{eqnarray}
\mathscr{D}_{grav}[N^a]:=\int_{\Sigma}d^3x N^a\Big[(\partial_a
A^j_b-\partial_b A^j_a)E^b_j-A^j_a\partial_bE^b_j\Big],
\end{eqnarray}
and the matter contribution is
\begin{eqnarray}
\mathscr{D}_{\rm matter}[N^a]:=\int_{\Sigma}d^3x N^a \pi
\partial_a\varphi.
\end{eqnarray}

In order to obtain the perturbed Hamiltonian density and the
perturbed diffeomorphism constraint up to the third order, we need
the following two relations. We expand $(\det E)^{\frac{1}{2}}$ and
$(\det E)^{-\frac{1}{2}}$ to the third order as follows.
\begin{eqnarray}
\quad\quad(\det
E)^{\frac{1}{2}}=\bar{p}^{\frac{3}{2}}\Big[&&1+\frac{1}{2\bar{p}}\delta^i_a\delta
E^a_i+\frac{1}{8\bar{p}^2}(\delta^i_a\delta
E^a_i)^2-\frac{1}{4\bar{p}^2}\delta^i_b\delta^j_a\delta E^a_i\delta
E^b_j+\frac{1}{48\bar{p}^3}(\delta^k_c\delta
E^c_k)^3\quad\quad\nonumber\\
&-&\frac{1}{8\bar{p}^3}(\delta^k_c\delta
E^c_k)(\delta^i_b\delta^j_a\delta E^a_i\delta
E^b_j)+\frac{1}{12\bar{p}^3}\delta^i_c\delta^k_b\delta^j_a\delta^k_a\delta
E^a_i\delta E^b_j\delta E^c_k\nonumber\\
&+&\frac{1}{12\bar{p}^3}\delta^i_b\delta^j_c\delta E^a_i\delta
E^b_j\delta E^c_k+...\Big],
\end{eqnarray}
and
\begin{eqnarray}
(\det
E)^{-\frac{1}{2}}=\bar{p}^{-\frac{3}{2}}\Big[&&1-\frac{1}{2\bar{p}}\delta^i_a\delta
E^a_i+\frac{1}{8\bar{p}^2}(\delta^i_a\delta
E^a_i)^2+\frac{1}{4\bar{p}^2}\delta^i_b\delta^j_a\delta E^a_i\delta
E^b_j-\frac{1}{48\bar{p}^3}(\delta^i_a\delta
E^a_i)^3\nonumber\\
&-&\frac{1}{8\bar{p}^3}(\delta^k_c\delta
E^c_k)(\delta^i_b\delta_a^j\delta
E^a_i\delta E^b_j)-\frac{1}{12\bar{p}^3}\delta^i_c\delta^k_b\delta^j_a\delta
E^a_i\delta E^b_j\delta
E^c_k\nonumber\\
&-&\frac{1}{12\bar{p}^3}\delta^i_b\delta^j_c\delta^k_a\delta
E^a_i\delta E^b_j\delta E^c_k+...\Big].
\end{eqnarray}

Thus the third order gravitational Hamiltonian density can be
written as
\begin{eqnarray}
\mathfrak{H}_{\rm
grav}^{(3)}=\mathfrak{H}_1^{(3)}+\mathfrak{H}_2^{(3)}+\mathfrak{H}_3^{(3)}
\end{eqnarray}
where
\begin{eqnarray}
\mathfrak{H}_1^{(3)}&:=&\Big[\epsilon_i^{jk}\frac{E^c_jE^d_k}{\sqrt{\det
E}}2\partial_c \Gamma^i_d\Big]^{(3)}\nonumber\\
&=&2\epsilon_i^{jk}\frac{\delta E^c_j\delta E^d_k}{\sqrt{\det
E}}\partial_c\delta \Gamma^i_d+2\epsilon_{i}^{jk}\delta
E^c_j\bar{E}^d_k[(\det E)^{-\frac{1}{2}}]^{(1)}\partial_c\delta
\Gamma^i_d\nonumber\\
&&+2\epsilon_i^{jk}\bar{E}^c_j\delta
E^d_k[(\det E)^{-\frac{1}{2}}]^{(1)}\partial_c\delta\Gamma^i_d
+2\epsilon_i^{jk}\bar{E}^c_j\bar{E}^d_k[(\det
E)^{-\frac{1}{2}}]^{(2)}\partial_c\delta \Gamma^i_d\nonumber\\
&=&-\frac{1}{\bar{p}^{\frac{5}{2}}}\delta^{kl}\delta^j_d\delta E^d_l\delta
E^c_j(\partial_c\partial_e\delta
E^e_k)+\frac{1}{2\bar{p}^{\frac{5}{2}}}\delta^{kl}\delta E^c_l\delta ^m_a\delta
E^a_m(\partial_c\partial_e\delta
E^e_k)\nonumber\\
&&-\frac{1}{4\bar{p}^{\frac{5}{2}}}\delta^{cl}(\delta^m_a\delta
E^a_m)^2(\partial_c\partial_e\delta
E^e_l)+\frac{1}{2\bar{p}^{\frac{5}{2}}}\delta^{cl}\delta^m_b\delta
E^a_m\delta E^b_n(\partial_c\partial_e\delta E^e_l),
\end{eqnarray}

\begin{eqnarray}
\mathfrak{H}_2^{(3)}&:=&\Big[\epsilon_i^{jk}\frac{E^c_jE^d_k}{\sqrt{\det
E}}\epsilon^i_{mn}\Gamma^m_c\Gamma^n_d\Big]^{(3)}\nonumber\\
&=&\epsilon_i^{jk}\frac{\bar{E}_k^d\delta E^c_j}{\sqrt{\det E}}
\epsilon^{i}_{mn}\delta \Gamma^m_c\delta\Gamma^n_d+\epsilon_i^{jk}
\frac{\bar{E}^c_j\delta E^d_k}{\sqrt{\det E}}\epsilon^i_{mn}\delta\Gamma^m_c
\delta\Gamma^n_d\nonumber\\
&&+\epsilon_i^{jk}\bar{E}^c_j\bar{E}^d_k
[(\det E)^{-\frac{1}{2}}]^{(1)}\epsilon^{i}_{mn}\delta \Gamma^m_c\delta\Gamma^n_d
\nonumber\\
&=&\frac{1}{4\bar{p}^{\frac{5}{2}}}\delta^k_d\delta
E^d_k\delta^{ij}(\partial_c \delta E^c_j)(\partial_a\delta
E^a_i)-\frac{1}{2\bar{p}^{\frac{5}{2}}}\delta^i_c\delta^{kj}\delta
E^c_j(\partial_b \delta E^b_k)(\partial_a\delta E^a_i),
\end{eqnarray}

\begin{eqnarray}
\mathfrak{H}_3^{(3)}&:=&\Big[-\epsilon_i^{jk}\frac{E^c_jE^d_k}{\sqrt{\det
E}}\epsilon^i_{mn}K^m_cK^n_d\Big]^{(3)}\nonumber\\
&=&-\epsilon_i^{jk}\frac{\delta E^c_j\delta E^d_k}{\sqrt{\det
E}}\epsilon^i_{mn}\bar{K}^n_d\delta
K^m_c-\epsilon_i^{jk}\frac{\bar{E}^d_k\delta E^c_j}{\sqrt{\det
E}}\epsilon^i_{mn}\delta K^m_c\delta
K^n_d\nonumber\\
&&-\epsilon_i^{jk}\frac{\delta E^c_j\delta E^d_k}{\sqrt{\det
E}}\epsilon^i_{mn}\bar{K}^m_c\delta K^n_d
-\epsilon_i^{jk}\frac{\bar{E}^c_j\delta E^d_k}{\sqrt{\det
E}}\epsilon^i_{mn}\delta K^m_c\delta
K^n_d\nonumber\\
&&-\epsilon_i^{jk}\delta E^c_j\delta E^d_k[(\det
E)^{-\frac{1}{2}}]^{(1)}\epsilon^i_{mn}\bar{K}^m_c\bar{K}^n_d-
\epsilon_i^{jk}\bar{ E}^c_j\bar{E}^d_k[(\det
E)^{-\frac{1}{2}}]^{(1)}\epsilon^i_{mn}\delta K^m_c\delta
K^n_d\nonumber\\
&&-\epsilon_i^{jk}\delta E^c_j\bar{E}^d_k[(\det
E)^{-\frac{1}{2}}]^{(1)}\epsilon^i_{mn}\delta K^m_c\bar{K}^n_d
-\epsilon_i^{jk}\delta E^c_j\bar{E}^d_k[(\det
E)^{-\frac{1}{2}}]^{(1)}\epsilon^i_{mn}\bar{K}^m_c\delta
K^n_d\nonumber\\
&&-\epsilon_i^{jk}\bar{E}^c_j\delta E^d_k[(\det
E)^{-\frac{1}{2}}]^{(1)}\epsilon^i_{mn}\bar{K}^n_d\delta K^m_c
-\epsilon_{i}^{jk}\bar{E}^c_j\delta E^d_k[(\det
E)^{-\frac{1}{2}}]^{(1)}\epsilon^i_{mn}\bar{K}^m_c\delta
K^n_d\nonumber\\
&&-\epsilon_i^{jk}\bar{E}^d_k \delta E^c_j[(\det
E)^{-\frac{1}{2}}]^{(2)}\epsilon^i_{mn}\bar{K}^m_c\bar{K}^n_d-\epsilon_i^{jk}\bar{E}^c_j\delta
E^d_k[(\det
E)^{-\frac{1}{2}}]^{(2)}\epsilon^i_{mn}\bar{K}^m_c\bar{K}^n_d\nonumber\\
&&-\epsilon_i^{jk}\bar{E}^c_j\bar{E}^d_k[(\det
E)^{-\frac{1}{2}}]^{(2)}\epsilon^i_{mn}\delta
K^m_c\bar{K}^n_d-\epsilon_i^{jk}\bar{E}^c_j\bar{E}^d_k[(\det
E)^{-\frac{1}{2}}]^{(2)}\epsilon^i_{mn}\bar{K}^m_c\delta
K^n_d\nonumber\\
&&-\epsilon_i^{jk}\bar{E}^c_j\bar{E}^d_k[(\det
E)^{-\frac{1}{2}}]^{(3)}\epsilon^i_{mn}\bar{K}^m_c\bar{K}^n_d\nonumber\\
&=&-\frac{2\bar{k}}{\bar{p}^{\frac{3}{2}}}\delta^k_d\delta
E^d_k\delta E^c_j\delta
K^j_c+\frac{2\bar{k}}{\bar{p}^{\frac{3}{2}}}\delta^j_d\delta
E^c_j\delta E^d_k\delta
K^k_c-\frac{2}{\bar{p}^{\frac{1}{2}}}\delta^d_n\delta K^n_d\delta
E^c_j\delta
K^j_c\nonumber\\
&&+\frac{2}{\bar{p}^{\frac{1}{2}}}\delta^d_m\delta E^c_j\delta
K^m_c\delta K^j_d
+\frac{\bar{k}^2}{2\bar{p}^{\frac{5}{2}}}(\delta^j_c\delta
E^c_j)^3-\frac{\bar{k}^2}{2\bar{p}^{\frac{5}{2}}}\delta^j_d\delta^k_c\delta^l_a\delta
E^c_j\delta E^d_k\delta
E^a_l\nonumber\\
&&+\frac{1}{2\bar{p}^{\frac{1}{2}}}(\delta^c_m\delta
K^m_c)^2\delta^l_a\delta E^a_l
-\frac{1}{2\bar{p}^{\frac{1}{2}}}\delta^c_n\delta^d_m\delta
K^m_c\delta K^n_d\delta^l_a\delta E^a_l
+\frac{2\bar{k}}{\bar{p}^{\frac{3}{2}}}\delta E^c_j\delta
K^j_c\delta^l_a\delta
E^a_l\nonumber\\
&&+\frac{\bar{k}}{\bar{p}^{\frac{3}{2}}}(\delta^j_c\delta
E^c_j)^2(\delta^d_n\delta K^n_d)
-\frac{\bar{k}}{\bar{p}^{\frac{3}{2}}}\delta^j_n\delta^d_c\delta
E^c_j\delta K^n_d\delta^l_a\delta E^a_l
-\frac{1}{2}\frac{\bar{k}^2}{\bar{p}^{\frac{5}{2}}}(\delta^j_c\delta
E^c_j)^3\nonumber\\
&&-\frac{\bar{k}^2}{\bar{p}^{\frac{5}{2}}}(\delta^l_b\delta^i_a\delta
E^a_l\delta E^b_i)(\delta^j_c\delta
E^c_j)-\frac{\bar{k}}{2\bar{p}^{\frac{3}{2}}}(\delta^l_a\delta
E^a_l)^2(\delta^d_n\delta
K^n_d)-\frac{\bar{k}}{\bar{p}^{\frac{3}{2}}}(\delta^l_b\delta^j_a\delta
E^a_l\delta E^b_j)(\delta^d_n\delta K^n_d)\quad\quad\nonumber\\
&&+\frac{\bar{k}^2}{8\bar{p}^{\frac{5}{2}}}(\delta^i_a\delta
E^a_i)^3+\frac{3\bar{k}^2}{4\bar{p}^{\frac{5}{2}}}(\delta^j_b\delta
E^b_j)(\delta^i_c\delta^k_a\delta E^a_i\delta
E^c_k)\nonumber\\
&&+\frac{\bar{k}^2}{2\bar{p}^{\frac{5}{2}}}\delta^i_c\delta^k_b\delta^j_a\delta
E^a_i\delta E^b_j\delta
E^c_k+\frac{\bar{k}^2}{2\bar{p}^{\frac{5}{2}}}\delta^i_b\delta^j_c\delta^k_a\delta
E^a_i\delta E^b_j\delta E^c_k\quad\quad\quad\quad.
\end{eqnarray}

The matter hamiltonian can be expressed as
\begin{eqnarray}
\quad\mathfrak{H}_{\pi}^{(3)}&=&\frac{1}{2}\delta\pi^2[(\det
E)^{-\frac{1}{2}}]^{(1)}+\bar{\pi}\delta \pi[(\det
E)^{-\frac{1}{2}}]^{(2)}+\frac{1}{2}\pi^2[(\det
E)^{-\frac{1}{2}}]^{(3)}\nonumber\\
&=&-\frac{1}{4\bar{p}^{\frac{5}{2}}}\delta^i_a\delta \pi^2\delta
E^a_i+\frac{\bar{\pi}}{\bar{p}^{\frac{3}{2}}}\delta\pi\Big[\frac{(\delta^i_a\delta
E^a_i)^2}{8\bar{p}^2}+\frac{\delta^i_b\delta^j_a\delta E^a_i\delta
E^b_j}{4\bar{p}^2}\Big]
+\frac{\bar{\pi}^2}{2\bar{p}^{\frac{3}{2}}}\cdot\nonumber\\
&\Big[-&\frac{(\delta^i_a\delta
E^a_i)^3}{48\bar{p}^3}-\frac{(\delta^i_c\delta^k_a\delta E^a_i\delta
E^c_k)(\delta^j_b\delta
E^b_j)}{8\bar{p}^3}-\frac{\delta^i_c\delta^k_b\delta^j_a\delta
E^a_i\delta E^b_j\delta
E^c_k}{12\bar{p}^3}-\frac{\delta^i_b\delta^j_c\delta^k_a\delta
E^a_i\delta E^b_j\delta E^c_k}{12\bar{p}^3}\Big],\quad\quad\quad
\end{eqnarray}
\begin{eqnarray}
\mathfrak{H}_{\rm \nabla}^{(3)}&=&\frac{\delta^{ij}\bar{E}^b_j\delta
E^a_i\partial_a\delta\varphi\partial_b\delta \varphi}{2\sqrt{\det
E}}+\frac{\delta^{ij}\bar{E}^a_i\delta
E^b_j\partial_a\delta\varphi\partial_b\delta\varphi}{2\sqrt{\det
E}}+\frac{\delta^{ij}\bar{E}^a_i\bar{E}^b_j}{2}[(\det
E)^{-\frac{1}{2}}]^{(1)}\partial_a\delta\varphi\partial_b\delta\varphi\nonumber\\
&=&\frac{\delta^{ai}\delta
E^b_i\partial_a\delta\varphi\partial_b\delta\varphi}{\bar{p}^{\frac{1}{2}}}-\frac{\delta^k_c\delta
E^c_k\delta^{ab}\partial_a\delta\varphi\partial_b\delta\varphi}{4\bar{p}^{\frac{1}{2}}},
\end{eqnarray}
\begin{eqnarray}
\quad\mathfrak{H}_{\rm \varphi}^{(3)}&=&\frac{1}{3!}\sqrt{\det
E}V_{,\varphi\varphi\varphi}(\bar{\varphi})\delta\varphi^3+\frac{1}{2}[(\det
E)^{\frac{1}{2}}]^{(1)}V_{,\varphi\varphi}(\bar{\varphi})\delta\varphi^2+[(\det
E)^{\frac{1}{2}}]^{(2)}V_{,\varphi}(\bar{\varphi})\nonumber\\
&&+\left[(\det
E)^{\frac{1}{2}}\right]^{(3)}V(\bar{\varphi})\nonumber\\
&=&\frac{1}{6}\bar{p}^{\frac{3}{2}}V_{,\varphi\varphi\varphi}(\bar{\varphi})\delta\varphi^3+\frac{1}{4}\bar{p}^{\frac{1}{2}}\delta^i_a\delta
E^a_iV_{,\varphi\varphi}(\bar{\varphi})\delta\varphi^2+\left[\frac{(\delta^i_a\delta
E^a_i)^2}{8\bar{p}^{\frac{1}{2}}}\right.\nonumber\\
&&\left.-\frac{\delta^i_b\delta^j_a\delta
E^a_i\delta
E^b_j}{4\bar{p}^{\frac{1}{2}}}\right]V_{,\varphi}(\bar{\varphi})\delta\varphi
+\bar{p}^{\frac{3}{2}}
\left[\frac{(\delta^k_c\delta
E^c_k)^3}{48\bar{p}^3}-\frac{(\delta^k_c\delta
E^c_k)(\delta^i_b\delta^j_a\delta E^a_i\delta
E^b_j)}{8\bar{p}^3}\right.\nonumber\\
&&\left.+\frac{\delta^i_c\delta^k_b\delta^j_a\delta E^a_i\delta
E^b_j\delta
E^c_k}{12\bar{p}^3}+\frac{\delta^i_b\delta^j_c\delta^k_a\delta
E^a_i\delta E^b_j\delta
E^c_k}{12\bar{p}^3}\right]V(\bar{\varphi}).\quad\quad\quad
\end{eqnarray}
Combined with the results obtained by \cite{bojowald09}, we get the
perturbed Hamiltonian as follows.
\begin{eqnarray}
\mathscr{H}_{\rm grav}^{(3)}[N]=\mathscr{H}_{\rm grav}^{(3)}[\delta
N]+\mathscr{H}_{\rm grav}^{(3)}[\bar{N}],
\end{eqnarray}
where
\begin{eqnarray}
\mathscr{H}_{\rm grav}^{(3)}[\bar{N}]=\frac{1}{2}\int_{\Sigma} d^3x
\bar{N}^{(3)}\mathfrak{H}_{{\rm grav}}^{(3)},
\end{eqnarray}
and
\begin{eqnarray}
\mathscr{H}_{\rm grav}^{(3)}[\delta N]=\frac{1}{2}\int_{\Sigma}d^3 x
\delta N \mathfrak{H}_{{\rm grav}}^{(2)},
\end{eqnarray}
and the matter Hamiltonian reads
\begin{eqnarray}
\mathscr{H}_{\rm matter}^{(3)}[\bar{N}]=\int_{\Sigma}d^3
x\bar{N}\Big[\mathfrak{H}_{\rm \pi}^{(3)}+\mathfrak{H}_{\rm
\nabla}^{(3)} +\mathfrak{H}_{\rm \varphi}^{(3)}\Big],
\end{eqnarray}
\begin{eqnarray}
\mathscr{H}_{\rm matter}^{(3)}[\delta N]=\int_{\Sigma}d^3x\delta N
\Big[\mathfrak{H}_{\rm \pi}^{(2)}+\mathfrak{H}_{\rm
\nabla}^{(2)}+\mathfrak{H}_{\rm \varphi}^{(2)}\Big].
\end{eqnarray}
Here we have ignored the high order correction terms caused by the
inverse volume, such as $\alpha^{(2)}H^{(2)}$, because in the in-in
formulism, these terms do not contribute to the non-Gaussianity.

The diffeomorphism constraint up to the third order is
\begin{eqnarray}
\mathscr{D}_{\rm grav}^{(3)}[\delta
N^a]&:=&\frac{1}{\gamma}\int_{\Sigma}d^3x\delta
N^a\Big[\frac{1}{2\bar{p}}\epsilon^{ij}_{b}(\partial_a\partial_c\delta
E^c_i)\delta E^b_j+\gamma\partial_a\delta K^j_b\delta
E^b_j-\frac{1}{2\bar{p}}\epsilon^{jk}_{a}(\partial_b\partial_c\delta
E^c_k)\delta E^b_j\nonumber\\
&-&\gamma\partial_b\delta K^j_a\delta
E^b_j-\frac{1}{2\bar{p}}\epsilon^{jk}_{a}\partial_c\delta
E^c_k\partial_b\delta E^b_j-\gamma\delta K^j_a\partial_b\delta
E^b_j\Big],
\end{eqnarray}
\begin{eqnarray}
\mathscr{D}_{\rm matter}^{(3)}[\delta N^a]:=\int_{\Sigma}d^3x \delta
N^a\delta\pi\partial_a\delta\varphi.
\end{eqnarray}

\section{Definitions of shape functions}
\label{shapefunc_definition}
For all the shapes considered in this paper $f(k)={\rm const.}$, so we have
\begin{equation}
S(k_1,k_2,k_3)=\mathscr S(\xi,y)\;.
\end{equation}
For the local shape
\begin{equation}
\label{shapefunc_local}
S_{{\rm local}}=\frac{k_1^3+k_2^3+k_3^3}{3k_1k_2k_3}\;,
\end{equation}
for the equilateral shape
\begin{equation}
\label{shapefunc_equil}
S_{{\rm equil}}=\frac{(k_1+k_2-k_3)(k_2+k_3-k_1)(k_3+k_1-k_2)}{k_1k_2k_3}\;,
\end{equation}
for the single shape
\begin{equation}
\label{shapefunc_shape}
S_{{\rm single}}=\frac{(3\epsilon_0-2\eta_0)\sum_ik_i^3+\epsilon_0\sum_{i\neq j}k_ik_j^2
+8\epsilon_0\sum_{i>j}k_i^2k_j^2/K}{k_1k_2k_3}\;,
\end{equation}
for the $\mathcal F_1$ shape
\begin{equation}
\label{shapefunc_F1}
S_{\mathcal F_1}=\left[\left(\frac{k_0}{k_1}\right)^{\sigma}+
\left(\frac{k_0}{k_2}\right)^{\sigma}+\left(\frac{k_0}{k_3}\right)^{\sigma}\right]\times S_{{\rm single}}\;,
\end{equation}
and for the $\mathcal F_2$ shape
\begin{equation}
\label{shapefunc_F2}
S_{\mathcal F_2}=\frac{\Big[2\omega_3'(\epsilon_0-\eta_0)+\omega_2'\epsilon_0\Big]
\sum_ik_i^3\left(k_0/k_i\right)^\sigma
+\omega_2'\epsilon_0\sum_{i\neq
j}k_j^2k_i\left(k_0/k_i\right)^{\sigma}}{k_1k_2k_3}\;.
\end{equation}


\end{document}